\documentclass[cup6b]{cupbook}
\usepackage{epsf,graphicx,amssymb}
\title[]
      {Long Gamma-Ray Burst Host Galaxies and their Environments}

\begin{document}
\setcounter{chapter}{12}

\author[J. P. U. Fynbo;  D. Malesani; P. Jakobsson]{Johan P. U. Fynbo$^1$, Daniele Malesani$^1$,
and P{\'a}ll Jakobsson$^2$\\
(1) Dark Cosmology Centre, Niels Bohr Institute, The University of Copenhagen,
Denmark\\
(2) Centre for Astrophysics and Cosmology, Science Institute, University of Iceland, Iceland}

\chapter{Long Gamma-Ray Burst Host Galaxies and their Environments}

\section{Introduction}

Host galaxies have played an important role in determining the nature of
Gamma-Ray Burst (GRB) progenitors even before the first optical afterglows were
detected.  After the first detections of host galaxies, their properties
provided strong evidence that long-duration GRBs were associated with massive
stellar deaths and hence the concept of using long-duration GRBs to probe the
evolution of the cosmic star-formation rate was conceived.

In this chapter we first briefly discuss some basic observational issues
related to what a GRB host galaxy is (whether they are operationally well
defined as a class) and sample completeness. We then describe some of the early
studies of GRB hosts starting with statistical studies of upper limits done
prior to the first detections, the first host detection after the
\textit{BeppoSAX} breakthrough and leading up to the current {\it Swift} era.
Finally, we discuss the status of efforts to construct a more complete sample
of GRBs based on {\it Swift} and end with an outlook. We only consider the host
galaxies of long-duration GRBs. For short GRBs we refer to Berger (2009). The
study of GRB host galaxies has previously been reviewed by van Paradijs et al.\
(2000) and Djorgovski et al.\ (2003).

\section{Early results based on GRB host galaxy studies}

\subsection{Pre-afterglow host galaxy studies}

Prior to the first afterglow detections a few gamma-ray bursts
with relatively small uncertainties on their derived position were studied
based on the Interplanetary Network (Hurley et al.\ 1993; see also Chapter 2).  Limits
on host galaxy magnitudes in such boxes coupled with distance estimates based
on $\log{N}$ vs. $\log{S}$ arguments (see also Chapter 3) suggested that GRB host galaxies
predominantly had to be subluminous (Fenimore et al.\ 1993). However, such
limits depended strongly on the assumed GRB luminosity function. As argued by
Woods \& Loeb (1995), above a certain width of the GRB luminosity function the
probability of detection of the then possibly very distant GRBs with apparently
faint hosts would be considerable (see also Larson 1997 and Wijers et al.\
1998). In fact both explanations turned out to contain part of the truth.

\subsection{The first host galaxy detections}

Obviously, the detection of the first optical afterglows (van Paradijs et al.\
1997) fundamentally changed the study of host galaxies. The first detection of
an extended source at the position of a GRB afterglow was for GRB\,970228 (Sahu
et al.\ 1997) -- the first GRB with a detected X-ray and optical afterglow
(Costa et al.\ 1997, van Paradijs et al.\
1997). At the time, it was not firmly proven that this extended source actually was the
host galaxy so the distance scale of GRBs was yet to be established. The next
GRB with a detected optical afterglow was GRB\,970508. This burst was found to have
a redshift of 0.695 (Metzger et al.\ 1997) and hence 
the extragalactic nature of (the majority of) long-duration GRBs was
established. GRB hosts were subsequently soon found to be
predominantly blue, star-forming galaxies, suggesting a young population origin
for the bursts (Paczy{\'n}ski 1998, Hogg et al. 1999, Christensen et al.\ 2004).  Another important point
was clear after the detection of GRB\,970508, namely that GRBs allow the
detection of distant (star-forming) dwarf galaxies that are very difficult to
detect with other methods (Natarajan et al.\ 1997). 

A few months later
relatively deep, early limits were obtained on the magnitude of the optical afterglow of
GRB\,970828. The non-detection of this optical afterglow suggested that some
GRBs occur along sightlines with substantial dust extinction in the observed
optical band (Groot et al. 1998). The third\footnote{Much later, in 2003, it
was established that an optical afterglow was also detected for GRB\,970815
(Soderberg et al. 2004).} GRB to have its optical afterglow detected was
GRB\,971214 for which a redshift of $z=3.42$ was established from the likely
host galaxy (Kulkarni et al.\ 1998). Hence, it was immediately clear that GRBs
allow us to probe ongoing star-formation throughout the observable Universe
(e.g., Wijers et al.\ 1998). 

In late March 1998 two further optical afterglows
were detected and then in April, GRB\,980425 was found to be associated with a
hyperluminous type Ic SN in a nearby dwarf galaxy at redshift $z=0.0085$
(Galama et al.\ 1998). The intrinsic fluence of this burst was
about 4 orders of magnitude fainter than the GRBs with optical afterglows
studied before and hence this discovery started discussions both on
low-luminosity bursts and the issue of chance projection. It is remarkable that 
roughly within a year after the first
detected optical afterglow some of the most important conclusions were already
reached: GRBs are related to massive stellar deaths, are located predominantly
in star-forming dwarf galaxies, are detected throughout most of the observable
Universe, are sometimes hidden by dust, and there seems to be a
population of low-luminosity GRBs only detectable in the relatively 
local Universe.

\section{Operational issues related to GRB host galaxy studies}

After having established that (at least the majority of) GRBs have hosts we
make a short interlude discussing operational issues related to GRB host
galaxies as a class.  It is prudent to keep these points in mind whenever
considering conclusions made about GRB hosts and their relation to other
classes of in particular high-redshift galaxies.

\subsection{Dark bursts and incomplete samples}

A crucial issue when discussing the nature of GRB host galaxies and the
implication thereof on the nature of GRB progenitors is sample completeness.
The detection of the GRB itself is of course limited by the sensitivity of the
gamma-ray detector and the GRB sample from a given mission will hence be
representative of a smaller and smaller part of the (possibly evolving) GRB
fluence distribution as one moves to higher and higher redshifts. However, in
terms of observational selection bias, the
GRB detection should not be affected by
host galaxy properties. In contrast, detection of the longer
wavelength afterglow emission, which is crucial for obtaining the precise
localization as well as measuring redshifts (see, e.g., Fiore et al.\ 2007), is
strongly dependent on the dust column density along the line-of-sight in the
host galaxy.

In the samples of GRBs detected with satellites prior to the currently
operating {\it Swift} satellite the fraction of GRBs with detected optical
afterglows was only about 30\% (Fynbo et al.\ 2001, Lazzati et al.\ 2002).
Much of this incompleteness was caused by random factors affecting ground-based
optical observations, such as weather or
unfortunate celestial positions of the bursts, but some remained undetected
despite both early and deep observations. It is possible that some of these so called
``dark bursts'' could be caused by GRBs in very dusty environments (Groot et
al.\ 1998) and hence the sample of GRBs with detected optical afterglows could
very well be systematically biased against dust obscured hosts (see also
Jakobsson et al.\ 2004a, Rol et al.\ 2005, 2007, Jaunsen et al.\
2008, Tanvir et al.\ 2008 for more recent discussions of the dark bursts).
Other causes for having a very faint optical afterglow are high redshifts (e.g. Greiner
et al.\ 2009, Ruiz-Velasco et al. 2007) or -- but only to some extent -- intrinsically hard spectra
(e.g., Pedersen et al.\ 2006).
In any case, such a high incompleteness imposes a large uncertainty on
statistical studies based on GRB host galaxies derived from these early
missions. 

It should be stressed that the conclusions based on these samples may
only be relevant for a minority of all GRBs and consequently a biased subsample
of the GRB host population. Galaxies hosting GRBs located in high-metallicity and hence
more dusty environments, (and we know already that such systems exist), will be
systematically underrepresented. Due to the much more precise and rapid X-ray
localization capability of {\it Swift} it is possible to build a much more
complete sample of GRBs from this mission as we will discuss later in this
chapter.

\subsection{Contamination from chance projection and Galactic transients}

\begin{figure}
\begin{center}
\epsfxsize=7cm \epsfbox{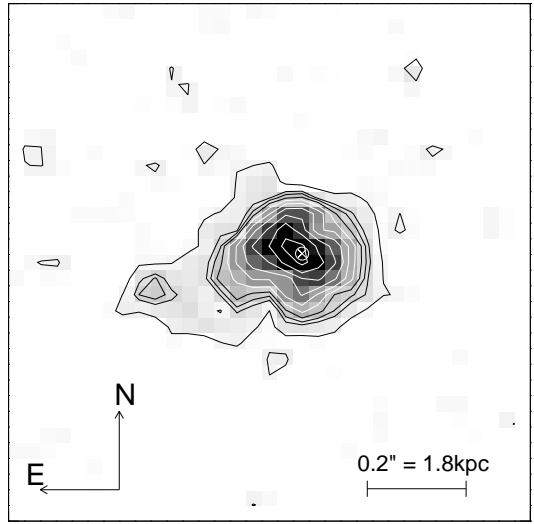}
\caption{
The {\it HST} 1$\times$1 arcsec$^2$ field around the host galaxy of GRB\,021004 at $z=2.33$ found with {\it HETE-2} (from Fynbo et al.\ 
2005). The GRB went off near the center of the galaxy. The position
of the GRB is marked with a cross and an error circle and in coincides with the
centroid of the galaxy to within a few hundredths of an arcsec.  In cases like
this there is no problem in identifying the correct host galaxy. However, in
cases of bursts localized to only a few arcsec accuracy, chance projection
needs to be considered.
}
\label{021004host}
\end{center}
\end{figure}

An important question to ask is: are GRB host galaxies operationally
well-defined as a class? The answer may seem to be trivially ``yes'', but
reality is more complex.  If we define the host galaxy of a particular burst to
be the galaxy nearest to the line-of-sight, we need to worry about chance
projection (Sahu et al.\ 1997, Band \& Hartmann 1998, Campisi \& Li 2008, Cobb \& Baylin 2008). In
the majority of cases where an optical afterglow has been detected and
localized with subarcsecond accuracy and where the field has been observed to
deep limits, a galaxy has been detected with an impact parameter less than 1
arcsec (see, e.g., Bloom et al.\ 2002, Fruchter et al.\ 2006 and
Fig.\,\ref{021004host} for an example). The probability for this to happen by
chance depends on the magnitude (and angular size) of the galaxy. The number of 
galaxies per
arcmin$^2$ has been well determined to deep limits in the various Hubble deep
fields.  To limits of $R=24$, 26 and 28 there are about 20, 80 and 400 galaxies
arcmin$^{-2}$. Hence, the probability to find an $R=24$ galaxy by chance in an
error circle with radius 0.5 arcsec is about 4$\times$10$^{-3}$. For an $R=28$
galaxy the probability is about 8\%. If the error circle is
defined only by the X-ray afterglow (with a radius of 2 arcsec in the best cases)
then we expect a random $R=24$ and $R=28$ galaxy in 6\% and all of the error
circles. For a sample of a few hundred GRBs chance projection should hence not
be a serious concern for GRBs localized to subsarcsecond precision, but for
error-circles with radius of a few arcseconds we expect many chance
projections. In some cases it may be possible to eliminate the chance
projections, e.g., based on conflicting redshift information from the afterglow
and proposed host (e.g., Jakobsson et al.\ 2004b); without such extra
information this is impossible.

Finally, it is worth noting that some events triggering the gamma-ray detectors
are Galactic high-energy transients rather than extragalactic GRBs. Examples
are GRB\,070610 (Kasliwal et al.\ 2008, Castro-Tirado et al. 2008) later renamed to SWIFT J195509.6+261406
and GRB\,060602B, which was found to be a low-mass X-ray binary (Wijnands et al.\
2009).
Judged from the high energy properties alone these bursts looked just
like GRBs, so in principle such events can contaminate GRB samples. Presumably,
most of these events will, like these two examples, be located at low
Galactic latitude and hence most can be rejected based on a Galactic latitude
cut in the sample selection.

\section{Status prior to the \textit{Swift} mission}
\subsection{GRBs as probes of star formation}
As outlined above, the properties of the first handful of detected hosts showed
a clear link to star-formation. Moreover, Mao \& Mo (1998) made a model
incorporating a power-law shaped luminosity function of GRBs and the assumption
that the GRB rate is proportional to the cosmic star-formation density. From
this model, Mao \& Mo (1998) were able to get remarkably good agreement with the
properties of observed hosts suggesting that GRBs were close to being good
tracers of star-formation. Hogg \& Fruchter (1999) reached a similar
conclusion.

\subsection{Biased tracers?}

\begin{figure}
\begin{center}
\epsfxsize=10cm \epsfbox{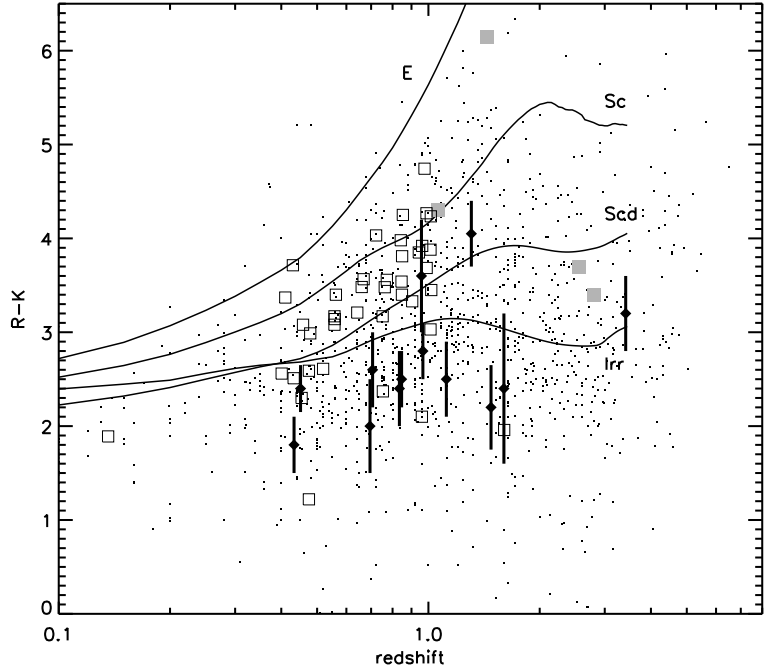}
\caption{
Observed $R-K$ colours versus redshift for the sample of GRB host galaxies
selected by Le Floc'h et al. (2003) ({\it filled diamonds}). The point of
the figure is that GRB host galaxies are bluer than other starburst galaxies
studied at similar redshifts. 
For comparison the authors also plot the
colours and redshifts for optically-selected field sources ({\it dots}) and of
ISO sources from the Hubble Deep Field ({\it open squares}) and of SCUBA
galaxies with confirmed redshifts ({\it filled squares}). See Le Floc'h et al.\
(2003) for more details and references. In addition, the authors found
that the $K$-band luminosities of GRB hosts were substantially fainter than,
e.g., the ISO galaxies which at redshifts around one are believed to 
dominate the total star-formation activity.
Solid curves
indicate the observed colours of local E, Sc, Scd and Irr galaxies if they were
moved back to increasing redshifts. 
}
\label{Emeric}
\end{center}
\end{figure}

\begin{figure}
\includegraphics[width=\textwidth]{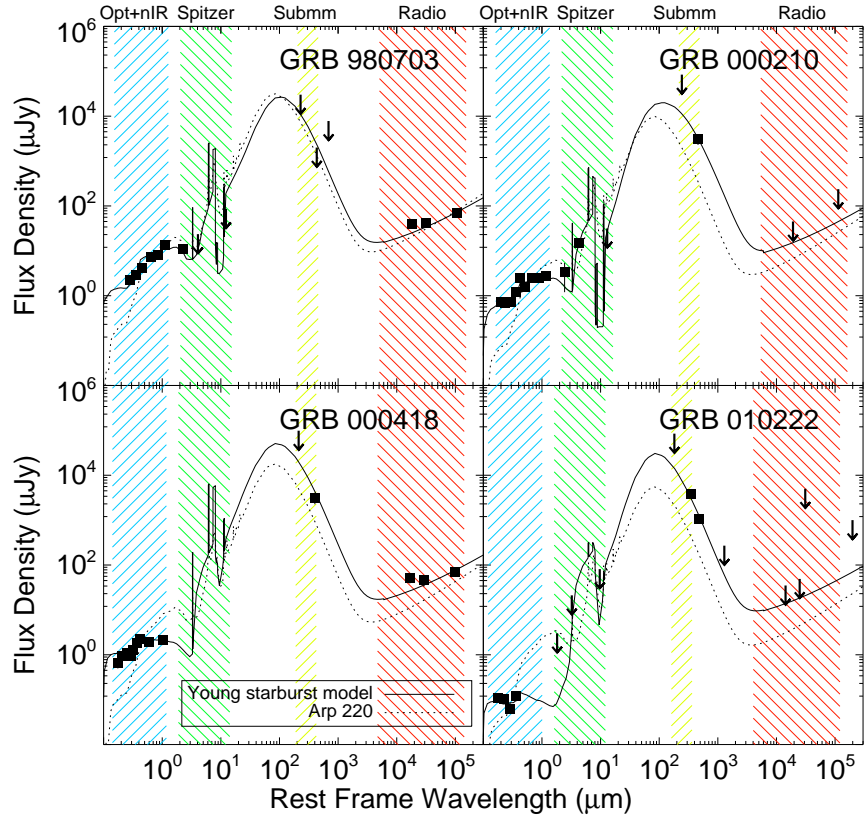}
\begin{center}
\vskip -0.2cm
\caption{Broad-band spectral energy distribution of four GRB hosts with firm
submillimetric or radio detections, demonstrating the role played by observations
in various wavebands (from Micha\l{}owski et al. 2008). Dotted lines show the
rescaled SED of the prototypical ultraluminous infrared galaxy Arp\,220,
while solid lines show synthetic best-fit models based on the GRASIL code
(Silva et al. 1998).\label{fg:broadband}}
\end{center}
\end{figure}

However, as the sample size grew evidence started to collect suggesting that
GRBs may be related only to massive stars with metallicity below a certain
threshold. As discussed in Chapter 10, such a metallicity dependence is expected
in the collapsar model. The first empirical evidence for this came with the
realization that the GRB hosts were fainter and bluer than expected according to
certain models about the nature of the galaxies dominating the integrated 
(over all galaxies)
star-formation rate density (Le Floc'h et al.\ 2003, see also Fig.~\ref{Emeric}).
Also, SCUBA imaging of GRB hosts in the sub-mm range produced only a few rather
tentative detections, again seemingly at odds with the expectations if GRB hosts
were selected in an unbiased way from all star-forming galaxies (Tanvir et al.\
2004). The analysis is complicated, and it has been pointed out by
Priddey et al.\ (2006) that ``there is sufficient uncertainty in models and
underlying assumptions, as yet poorly constrained by observation (e.g., the
adopted dust temperature) that a correlation between massive, dust-enshrouded
star formation and GRB production cannot be firmly ruled out.'' The issue of
dust temperature has been discussed in detail by Micha\l{}owski et al. (2008;
see also Fig.~\ref{fg:broadband}).
They find that the few GRB hosts that have been tentatively detected in the
sub-mm range have hotter dust and lower masses than typical sub-mm detected
galaxies.

Further circumstantial evidence for a preference towards low metallicity came
from the observation that Lyman-$\alpha$ (Ly$\alpha$) emission seemed to be
ubiquitous from GRBs hosts (Fynbo et al. 2003a, Jakobsson et al. 2005b). At this
point, around 2003, redshifts had been measured for ten $z>2$ GRBs.
%HST images of the three brightest of these are shown in Fig.~\ref{hosts}. As
%seen, these 
Ly$\alpha$ emission was detected for 5 of these and for the remaining 5 it was
not yet
searched for to sufficient depth to allow detection of even a large equivalent
width emission line. As only 25\% of continuum selected starbursts at
similar redshifts are Ly$\alpha$ emitters and as Ly$\alpha$ emission on
theoretical grounds should be more common for metal poor starbursts (Charlot \&
Fall 1993; but see also Mas-Hesse et al.\ 2003) this suggested that there could
be a low metallicity bias in making GRBs. However, the reason could also be an
observational bias against dusty (and hence likely higher metallicity) GRBs or
simply that the majority of the star formation is associated with
low-luminosity galaxies, which tend to be metal poor in accordance with the
luminosity-metallicity relation. The broad-band luminosity distribution of
$z>2$ GRB hosts was found by Jakobsson et al. (2005b) to be consistent with the
assumption that GRBs are selected from the rest-frame UV selection function
based on the total UV emission per luminosity bin. Assuming that the rest-frame
UV emission is proportional to the star formation rate this suggested that GRBs
cannot be strongly biased towards low metallicity.

%\begin{figure}[t]
%\begin{center}
%\includegraphics[width=0.379\textwidth]{../fynboj3.ps}
%\includegraphics[width=0.288\textwidth]{../fynboj4.ps}
%\includegraphics[width=0.288\textwidth]{../fynboj5.ps}
%\vskip -0.2cm
%\caption{
%HST images of the host galaxies of GRBs 000926 ($z=2.04$, $R=24.0$ AB), 011211
%($z=2.14$, $R=25.0$ AB), and 021004 ($z=2.33$, $R=24.4$ AB). The contours
%show the morphology of Ly$\alpha$ emission, as measured from ground-based
%narrow band imaging (Fynbo et al.\ 2002; Fynbo et al.\ 2003; Jakobsson et al.\
%2005).  GRB~000926 occurred near the centre of the right-most knot. The
%locations of GRB~011211 and GRB~021004 are marked with crosses (for GRB~021004
%see also Fig.~\ref{021004host}). The field sizes
%for the GRB~011211 and GRB~021004 images are the same.
%}
%\label{hosts}
%\end{center}
%\end{figure}

\begin{figure}[h]
\begin{center}
\includegraphics[width=0.99\textwidth]{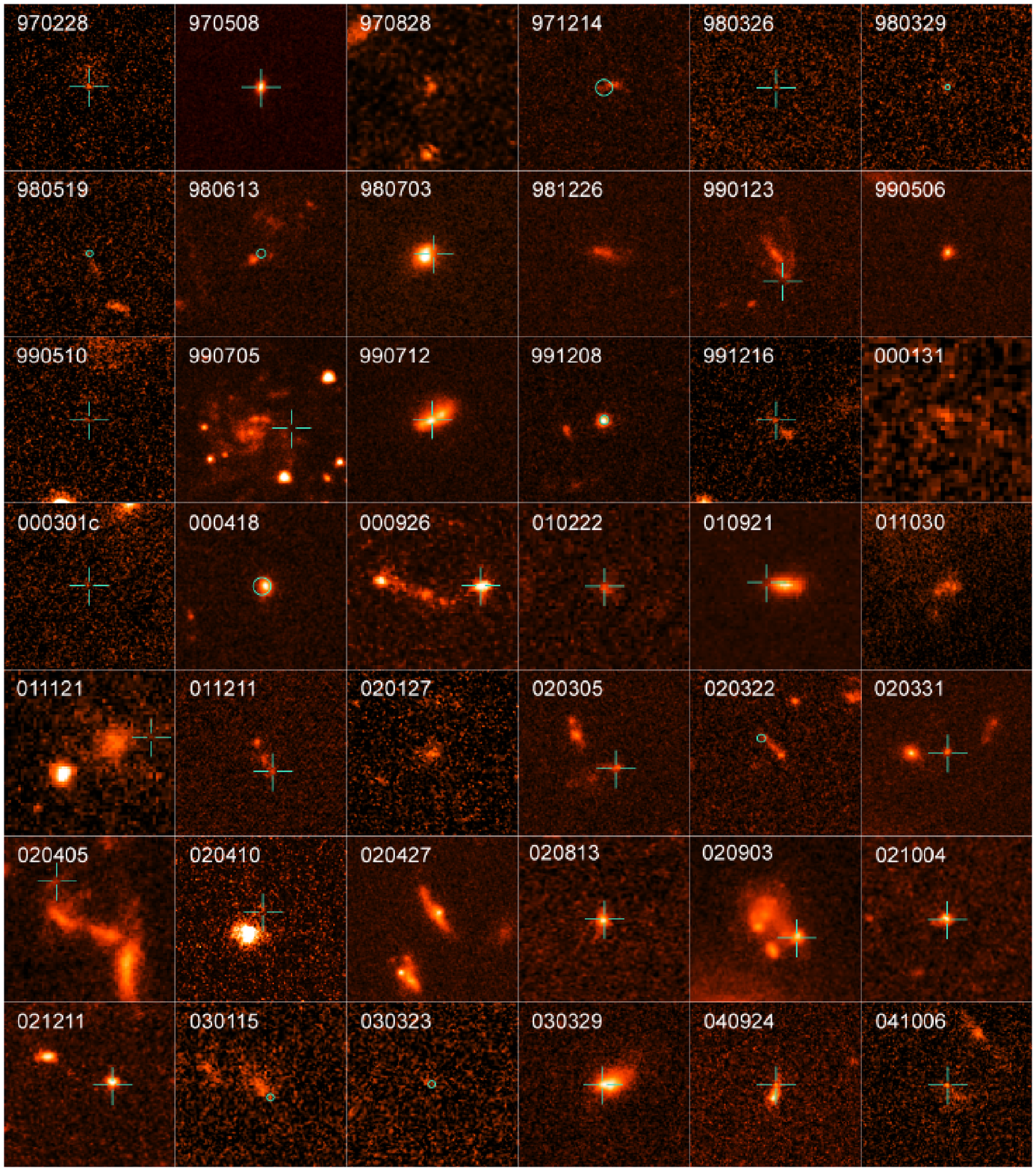}
\vskip -0.2cm
\caption{
A mosaic of GRB host galaxies imaged with {\it HST} (from Fruchter et al.\ 2006). Each
individual image corresponds to a square region on the sky $3.75$ arcsec on a
side. These images were taken with the {\it HST}. In cases where the location of the GRB on the host
is known to better than $0.15$ arcsec, the position of the GRB is shown by a green
mark. If the positional error is smaller than the point spread function of the
image ($0.07$ arcsec for STIS and ACS, $0.13$ arcsec for WFPC2) the position is
marked by a cross-hair, otherwise the positional error is indicated by a
circle. 
Due to the redshifts of the hosts, these images generally correspond
to blue or ultra-violet images of the hosts in their rest frame, and thus
detect light largely produced by the massive stars in the hosts.
}
\label{andyhosts}
\end{center}
\end{figure}

A very important result is that GRBs and core-collapse (CC) SNe are found in
different environments (Fruchter et al.\ 2006 and Fig.~\ref{andyhosts}).  GRBs
are significantly more concentrated on the very brightest regions of their host galaxies
than CC SNe.  The same study also found that GRB host galaxies at
$z<1$ are fainter and more irregular than the host galaxies of CC
SNe.  Fruchter et al.\ (2006) suggest that these results may imply that long-duration
GRBs are associated with the most extremely massive stars and may be restricted
to galaxies of limited chemical evolution. This would directly imply that
long-duration GRBs are relatively rare in galaxies such as our own Milky Way.

This study is also based on incomplete pre-{\it Swift} samples, but as the SN
samples are, if anything, more biased against dusty regions than GRBs, the
fact that GRB hosts have lower luminosities than CC SN hosts
does seem to provide substantial evidence that GRBs are biased towards
massive stars with relatively low metallicity. Regarding the size of the
effect, Wolf \& Podsiadlowski (2007) find, based on an analysis of the Fruchter
et al.\ (2006) data, that the metallicity threshold cannot be significantly
below half the solar metallicity.  Concerning the different environments of
CC SNe and GRBs it has recently been found that type Ic SNe have
similar positions relative to their host galaxy light profiles as GRBs,
whereas all other SN types have (similar) distributions less centred on their
host light than GRBs and SN Ic (Kelly et al.\ 2007). 

Larsson et al.\ (2007),
modeling the distribution of young star clusters and total light
in the local starburst NGC 4038/39 (the Antennae),
find that the different distributions of GRBs and CC SNe relative 
to their
host light can be naturally explained by assuming different mass ranges for the
typical progenitor stars: $>8~M_{\odot}$ for typical CC SNe and
$>20~M_{\odot}$ for GRB progenitors. The picture is complicated by the
finding that type Ic SNe typically are found in substantially more metal rich
environments than GRBs (Modjaz et al.\  2008).  It is well established that
WR stars become more frequent with {\it increasing} metallicity - opposite to
GRBs that, if anything, are biased towards low metallicity. Taken together, these
results suggest that progenitors of GRBs and ``normal'' type Ic SNe are two
different subsets of the $>20~M_{\odot}$ stars. For a thorough discussion of
the relation between WR stars, SN Ic and GRBs we refer to Crowther (2007). In
conclusion, there seems to be a low metallicity preference for GRBs, but the
metallicity threshold cannot be much below half solar and we need a more
unbiased and uniform sample to estimate the severity of the effect (see also Chapter 10).

\subsection{The galactic environment of GRB hosts}

The galactic environments of GRBs have so far not been studied much. At low
redshifts Foley et al.\ (2006) studied the field of the GRB\,980425 
host galaxy,
which was reported to be a member of a group. However, based on
redshift measurements of the proposed group members, Foley et al.\ (2006) could
establish that the host of GRB\,980425 is an isolated dwarf
galaxy\footnote{H$\alpha$ imaging of the host has revealed a very faint
companion about 1 arcmin NE of the host (L. Christensen, private
communication).}. Levan et al.\ (2006) also proposed GRB\,030115 to be
connected to a cluster around $z \sim 2.5$ based on photometric redshifts. At
redshifts $z\gtrsim2$ a few GRB fields have been studied using narrow band
Ly$\alpha$ imaging (Fynbo et al.\ 2002, Fynbo et al.\ 2003a, Jakobsson et al.\
2005b). In all cases several other galaxies at the same redshift as the GRB host
were identified, but it is not certain whether the galaxy densities in these
fields are higher than in blank fields as no blank field studies have been
carried out at similar redshifts.  However, the density of Ly$\alpha$ emitters
was found to be as high as in the fields around powerful radio sources that
have been proposed to be forming proto-clusters (Kurk et al.\ 2000), which
would suggest that GRBs could reside in overdense fields at $z\gtrsim$ 2.
Bornancini et al.\ (2004), on the other hand, argue for a low galaxy density in
GRB host galaxy environments. In conclusion, the evidence is currently too
sparse to establish whether GRB hosts are located in special environments.

\subsection{GRB host absorption line studies}

GRB host galaxies have the unique advantage over galaxies selected on the
base of their emission that spectroscopy of their afterglows (X-ray, optical,
and other bands) can reveal detailed information about the gas in the host
galaxy ranging from the circumstellar material to halo gas. %The field of
%GRB afterglow spectroscopy is currently very active and rapidly evolving and 
%here we can only point out some of the main conclusions. 

\begin{figure}[h]
\begin{center}
\includegraphics[width=1.00\textwidth]{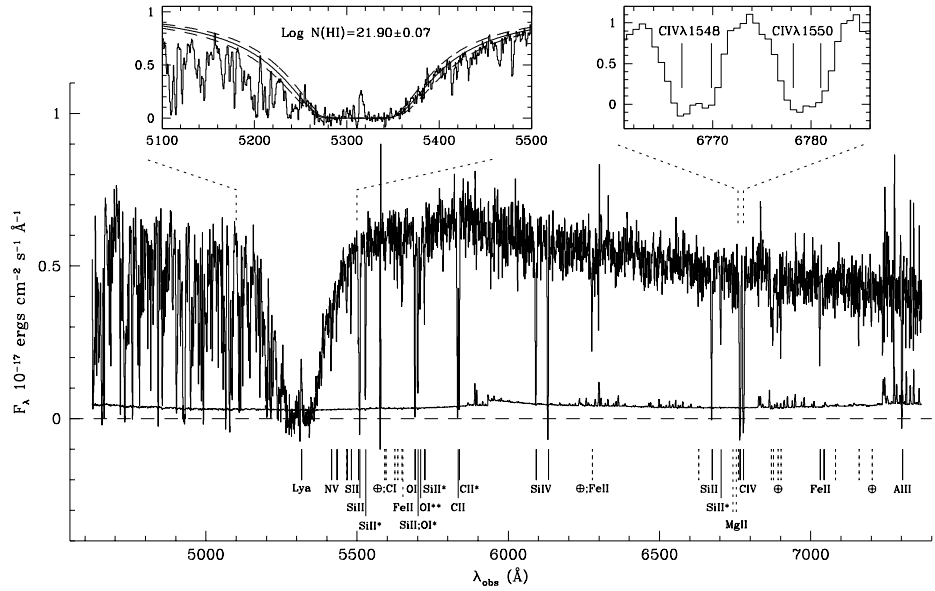}
\caption{
The spectrum of the optical afterglow of GRB\,030323 (from Vreeswijk et 
al.\ 2004). The insets show close-ups on the DLA (with Ly$\alpha$ in emission) and of the C IV doublet, showing the presence of at least two distinct
components.
}
\label{030323}
\end{center}
\end{figure}

Two interesting examples of what GRB afterglow spectroscopy can teach us about
GRB hosts are the cases of GRB\,030323 at redshift $z=3.372$ (Vreeswijk et al.\
2004) and GRB\,080607 at $z=3.036$ (Prochaska et al.\ 2009). As seen in
Fig.~\ref{andyhosts} the host galaxy of GRB\,030323 is among the faintest
detected (it has a magnitude of about $R=28$). The spectrum of the optical
afterglow of the event is shown in Fig.~\ref{030323}.  The dominant feature in
the spectrum is a very large damped Ly$\alpha$ absorption (DLA) line
corresponding to a redshift of $z=3.37$ (shown in more detail in the inset in
the upper left corner). A DLA is a hydrogen absorption line with column density 
above $2\times10^{20}$ cm$^{-2}$, where the total equivalent width is dominated by
the broad damping wings. Also seen are numerous low- and high-ionization lines
at a redshift $z=3.3718\pm0.0005$. The inferred neutral hydrogen column
density, $\log{(N_{\rm HI}/{\rm cm}^{-2})} = 21.90\pm0.07$, is very large -- higher than
in DLAs seen against the light of QSOs (Wolfe et al.\ 2005). From an analysis of 
the metal line
strengths compared to the hydrogen column density, a sulphur abundance of
about 0.05 solar is determined. %, while the metallicity of the gas as
%measured from sulphur is [S/H]=$-$1.26$\pm$0.20. 
An upper limit to the H$_2$
molecular fraction of 
$2N_{{\rm H}_2}/(2N_{{\rm H}_2}+N_{\rm HI}) < 10^{-6}$ 
is derived from an
analysis of the absence of molecular lines. 

\begin{figure}[h]
\begin{center}
\includegraphics[width=0.75\textwidth,angle=90]{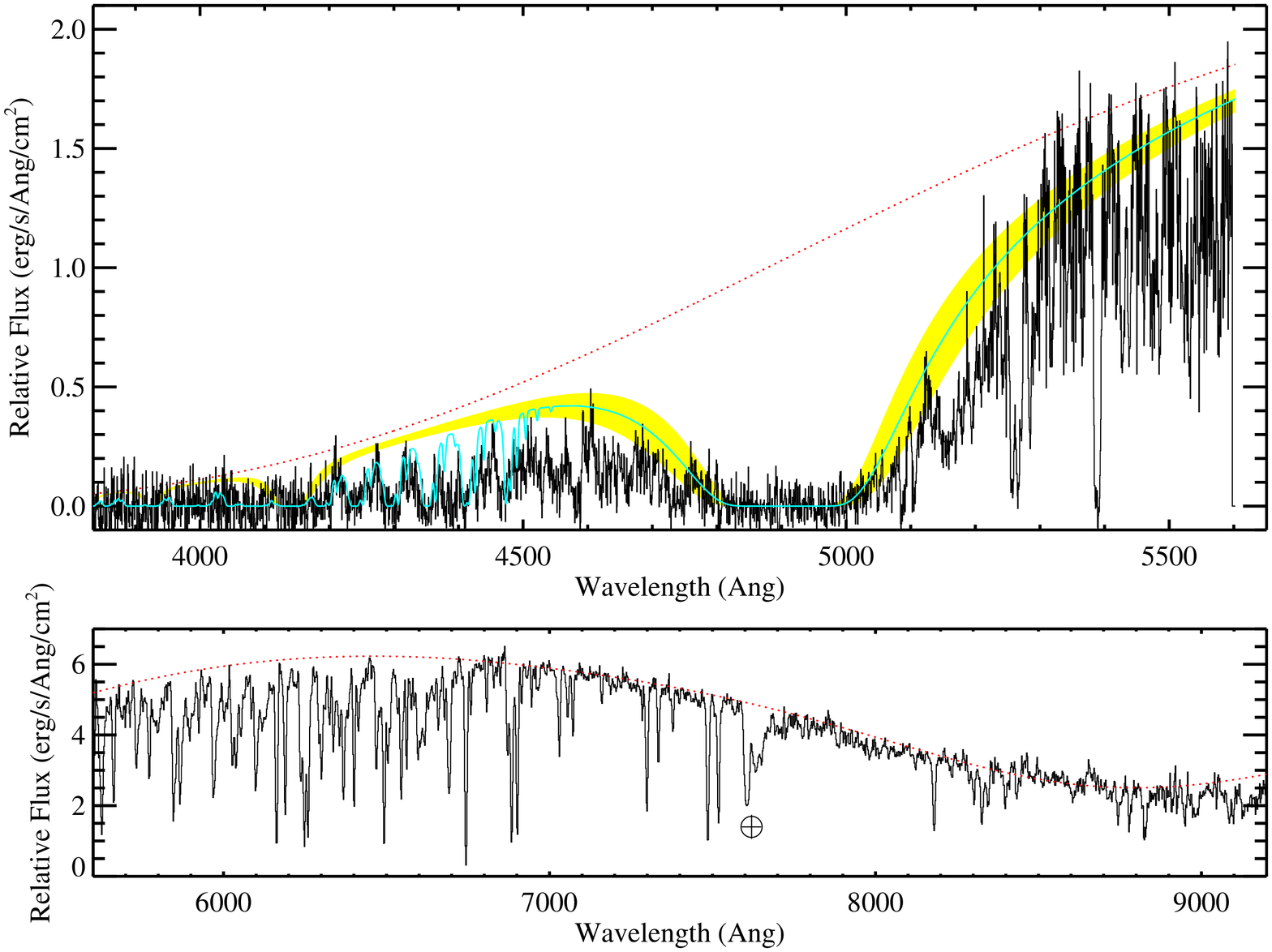}
\caption{
The spectrum of the optical afterglow of GRB\,080607 (from Prochaska et 
al.\ 2009). 
}
\label{fig:080607}
\end{center}
\end{figure}

In the DLA trough, a
Ly$\alpha$ emission line is detected, which corresponds to a star formation
rate (not corrected for dust extinction) of roughly $1~M_{\odot}$ yr$^{-1}$.
All these results are consistent with the host galaxy of GRB\,030323 containing
low metallicity gas with a low dust content. In addition, fine-structure lines 
of silicon, Si II*, are detected in the spectrum. These have not been clearly detected
in QSO-DLAs suggesting that these lines are produced in the vicinity of the GRB
explosion site. The optical spectrum of GRB\,080607 displays an even larger HI
column density of $\log{(N_{\rm HI}/{\rm cm}^{-2})} = 22.7$, but contrary to GRB\,030323, the
GRB\,080607 host is responsible for a very metal rich absorption system (close to solar
metallicity) with significant dust extinction and a clear detection of both
H$_2$ and CO molecular absorption (Fig.~\ref{fig:080607}). At the time of
writing (June 2010) the host galaxy has not been detected.

\begin{figure}[h]
\begin{center}
\includegraphics[width=4.0in,angle=90]{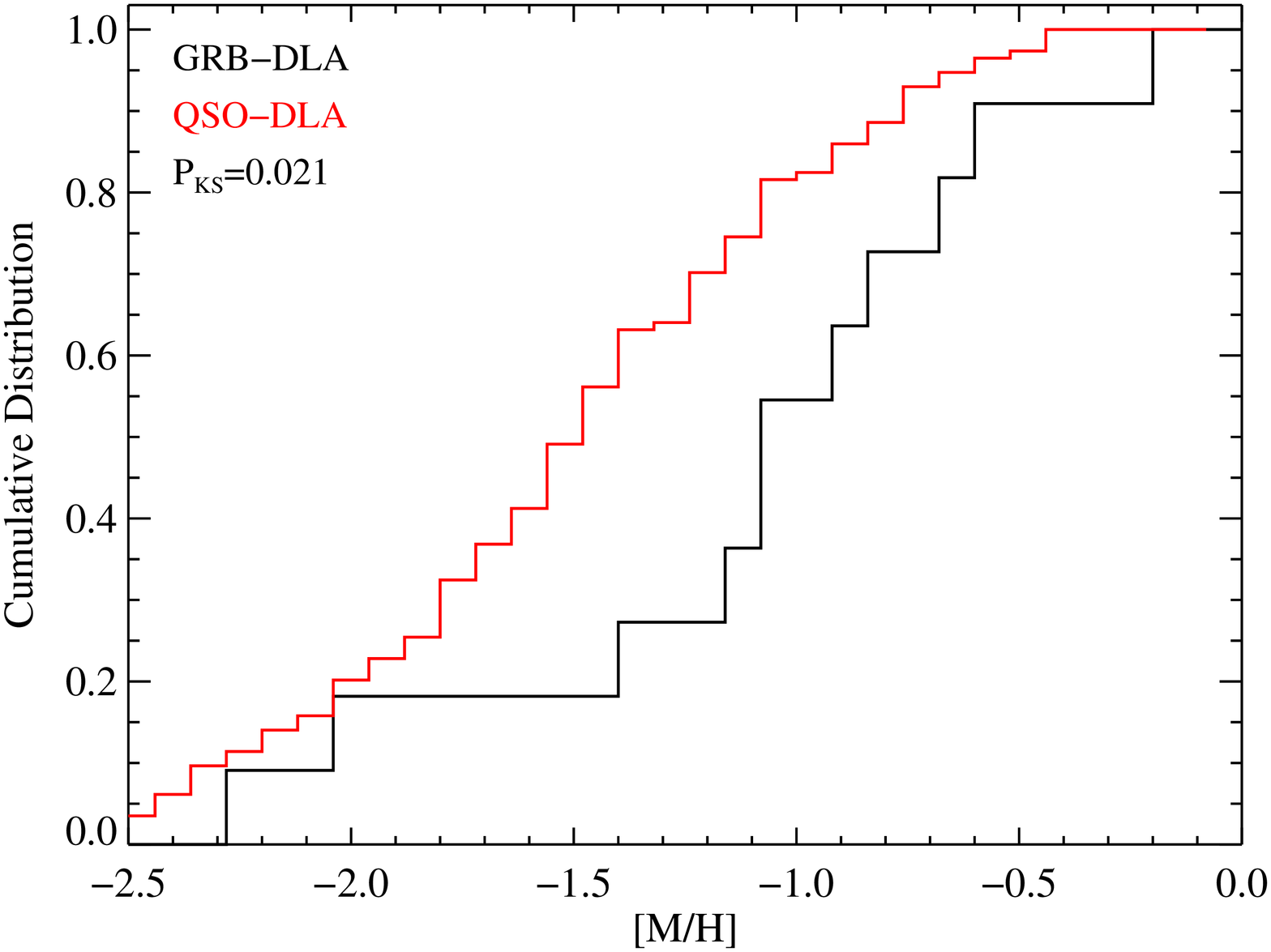} 
\caption{
The histograms show the cumulative distribution of QSO-DLA and GRB-DLA
metallicities in the statistical samples compiled by Prochaska et al.\ (2003) 
and
Prochaska et al.\ (2007). The GRB-DLA metallicities are systematically
higher than the QSO-DLA metallicities. Based on a KS test, the probability that the two observed distributions
are drawn from the same parent distribution is 2.1\% (see also Fynbo et al.\ 2008 for
a simple model of these two distributions).
}
\label{Z}
\end{center}
\end{figure}

Dozens of similar quality spectra have been obtained, some with
high resolution spectrographs (e.g., Chen et al.\ 2005, Starling et al.\ 2005,
Vreeswijk et al.\ 2007, D'Elia et al.\ 2007). The
properties of the GRB absorbers as a class are discussed in Savaglio (2006),
Fynbo et al.\ (2006) and Prochaska et al.\ (2007). The GRB absorbers are
characterized by HI column densities spanning five orders of magnitude from
$\sim$10$^{17}$ cm$^{-2}$ to $\sim$10$^{23}$ cm$^{-2}$ and metallicities
spanning two orders of magnitude from 1/100 to nearly solar (see Fig.~\ref{Z}).
Calura et al.\ (2009) have taken the first steps towards modelling the 
abundance ratios in GRB hosts using models for chemical evolution in galaxies. 
The majority of the GRB absorbers have metallicities exceeding the cosmic mean
metallicity of atomic gas at $z>2$ as determined from QSO-DLAs. The difference
in abundances between the QSO DLAs and GRB absorbers can be reconciled in a
simple model where the two populations are drawn randomly from the distribution
of star-forming galaxies according to their star formation rate and HI cross
section, respectively (Fynbo et al.\ 2008). However, it should be noted that
these results are based on a small sample of GRBs in which most have bright
optical afterglows.

\section{GRB host galaxies in the \textit{Swift} era}
The currently operating {\it Swift} satellite (Gehrels et al.\ 2004
and Chapter 5) has
revolutionized GRB research with its frequent, rapid, and precise localization
of both long and short duration GRBs. The breakthrough enabled by {\it Swift} was the ability to build a much more complete sample of localized GRBs and hence of GRB host galaxies. By complete we here mean unbiased in
terms of the optical properties of the GRB afterglows.

\subsection{Building a complete sample of \textit{Swift} GRBs}

The {\it Swift} satellite has been
superior to previous GRB missions due to the combination
of several factors: {\it i)} it detects GRBs at a rate of about two bursts per
week, about an order of magnitude larger than the previous successful \textit{BeppoSAX}
and {\it HETE-2} missions; {\it ii)} with its X-Ray Telescope (XRT) it localizes the
bursts with a precision of about 5 arcsec (often later refined to less than 2
arcsec), also orders of magnitude better than previous missions; {\it iii)} it
has a much shorter reaction time, allowing the study of the evolution of the
afterglows literally seconds after the burst, sometimes while the prompt
$\gamma$-ray emission is still being emitted. For a detailed description of the {\it Swift} era see also Chapter 5. 

We will here discuss the status of an ongoing effort to build such a complete
sample (see, e.g., Jakobsson et al.\ 2006a for earlier descriptions of this
work). Rather than including all {\it Swift} detected GRBs only
those GRB afterglows with favourable observing conditions are included, 
in particular those
fulfilling the following selection criteria:

\begin{enumerate}
\item{XRT afterglow detected within 12 hr;}
\item{small foreground Galactic extinction: $A_V<0.5$ mag;}
\item{favourable declination: $-70^\circ < \delta < +70^\circ$;}
\item{Sun-to-field angular distance larger than 55$^\circ$.}
\end{enumerate}

By introducing these constraints the sample is not biased towards GRBs with
optically bright afterglows, but will contain bursts for 
which useful follow-up observations are likely to be secured. 

The fact that 6--10 m class telescopes have made tremendous efforts to secure
redshifts means that this sample has a much higher redshift completion than for
pre-{\it Swift} samples (see Fig.~\ref{zdist}). Still, it is clear that we will not
get redshifts for all bursts from spectroscopy of the afterglows for multiple
reasons. In a small fraction of the cases where a spectrum of the afterglow is
secured no redshift can be measured. This is either due to 
lack of significantly detected absorption lines or
because the spectrum is simply too noisy. For these
bursts the only way to measure the redshift is via spectroscopy of the host
galaxy, but this is a challenging task due to the faintness of most GRB hosts
(see below).

\subsection{The redshift distribution of \textit{Swift} GRBs: current status}
\begin{figure}[h]
\begin{center}
\includegraphics[width=5.3in]{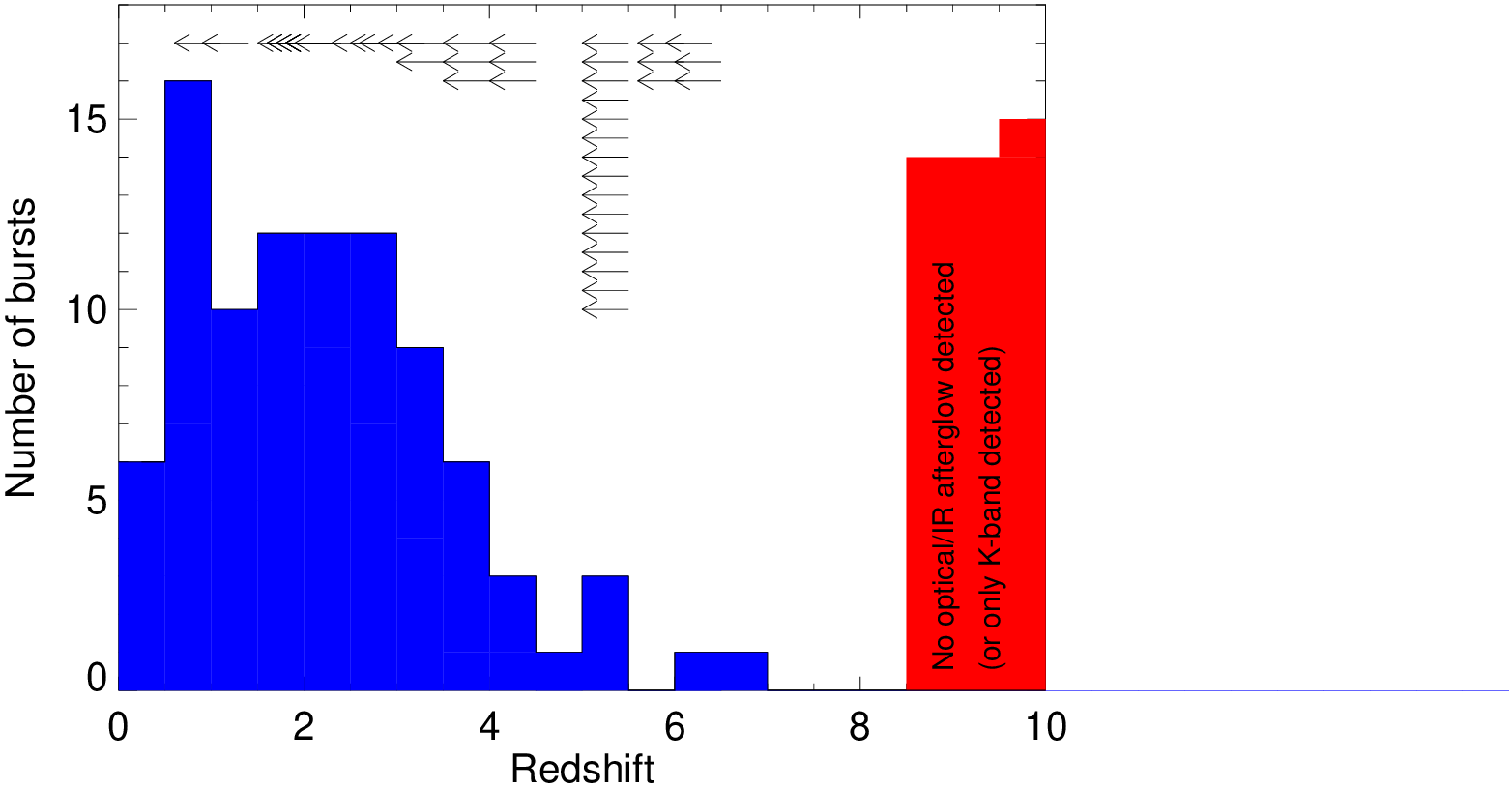} 
\caption{
Redshift distribution (up to March 2009) of {\it Swift} GRBs localized
with the X-ray telescope and with low foreground extinction $A_V\le0.5$. 
Bursts for which only an upper limit on the redshift could be established so
far are indicated by arrows. The histogram at
the right indicates the bursts for which no optical/$J$/$H$ afterglow was
detected and hence no redshift constraint could be inferred (see Ruiz-Velasco
et al.\ 2007 for a full discussion of an earlier version of this plot).
}
\label{zdist}
\end{center}
\end{figure}

The first conclusion from Fig.~\ref{zdist} is that most {\it Swift} GRBs are
very distant. {\it Swift} GRBs are more distant than GRBs from previous
missions due to the higher sensitivity of the satellite to the lower energies
prevalent in the more distant events (Fiore et al.\ 2007). The faster reaction
time of the satellite probably also contributes to shifting the median 
redshift upwards. The median and mean
redshift are now both around 2, while for previous missions it was closer to 1
(Jakobsson et al.\ 2006a).  The record holders are GRB\,080913 at $z=6.7$ 
(Greiner et al.\ 2009; Pattel et al.\ 2010) and GRB\,090423 at $z=8.2$
(Tanvir et al.\ 2010; Salvaterra et al.\ 2010).
It is striking how events at redshifts as large as $\sim 8$ can be detected within
such a small sample. For comparison, only a few QSOs are detected at $z \sim 6$
(and none at $z \sim 8$!) out of a sample of hundred thousand QSOs. 
Given the current redshift completeness level, the data are consistent with a
broad range of redshift distributions (see, e.g., Jakobsson et al.\ 2006a).
%Remarkably, the redshift
%distribution, measured for just over 50\% of all bursts, is consistent with the
%redshift distribution predicted if GRBs are unbiased tracers of star formation
%(see, e.g., Wijers et al.\ 1998, Jakobsson et al. 2006a and
%http://raunvis.hi.is/$\sim$pja/GRBsample.html for a regularly updated
%analysis). However, given that the redshift completeness is only slightly larger
%than 50\%, the data are consistent with a broad range of redshift distributions.

\subsection{HI column densities}
\begin{figure}
\begin{center}
\includegraphics[width=5in]{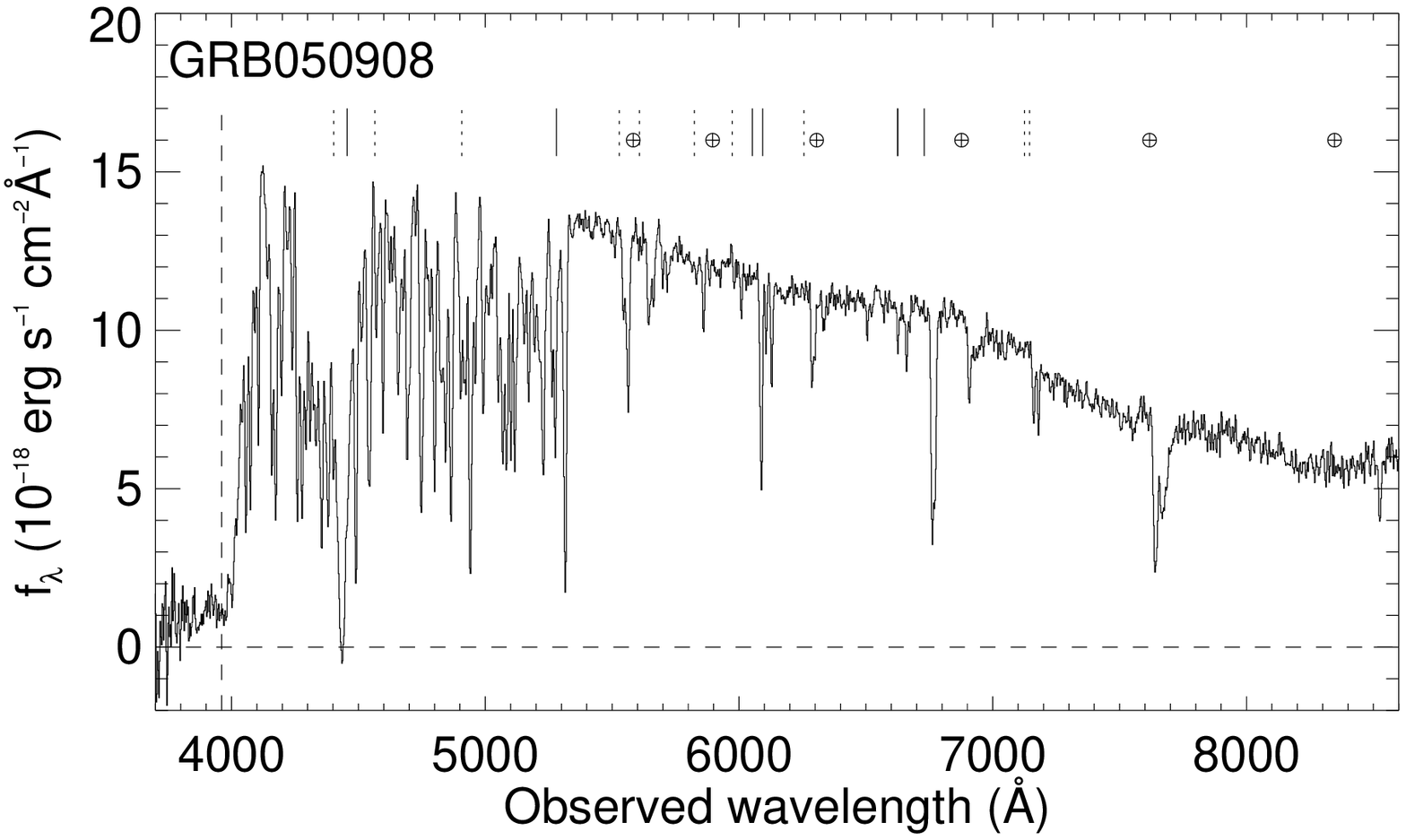}
\caption{
The {\it VLT}\/FORS2 spectrum of the afterglow of GRB\,050908 (Fynbo et al.\ 2009)
Plotted is the flux-calibrated 1-dimensional
spectrum against observed wavelength. The vertical
dashed line shows the position of the Lyman limit at the
redshift of the GRB ($z=3.343$). As seen, there
is clear excess flux blueward of the Lyman limit.
}
\label{LyC}
\end{center}
\end{figure}

The HI column density distribution for GRB sightlines is extremely broad. It
covers a range of about 5 orders of magnitude from $\sim10^{17}$ cm$^{-2}$
(Fig.~\ref{LyC}, Chen et al.\ 2007) to nearly $10^{23}$ cm$^{-2}$ 
(Jakobsson et al.\ 2006b). It still remains to be understood if this distribution
is representative of the intrinsic distribution of HI column densities
towards massive stars in galaxies or if the distribution is rather controlled
by the ionizing emission from the afterglows themselves. In any case, as 
pointed out by Chen et al. (2007), the HI column density distribution provides
an upper limit to the escape fraction of Lyman continuum emission from 
star-forming galaxies. This is crucial for the issue of determining the sources
for the ionising photons in the metagalactic UV background.

\begin{figure}
\begin{center}
\includegraphics[width=5in]{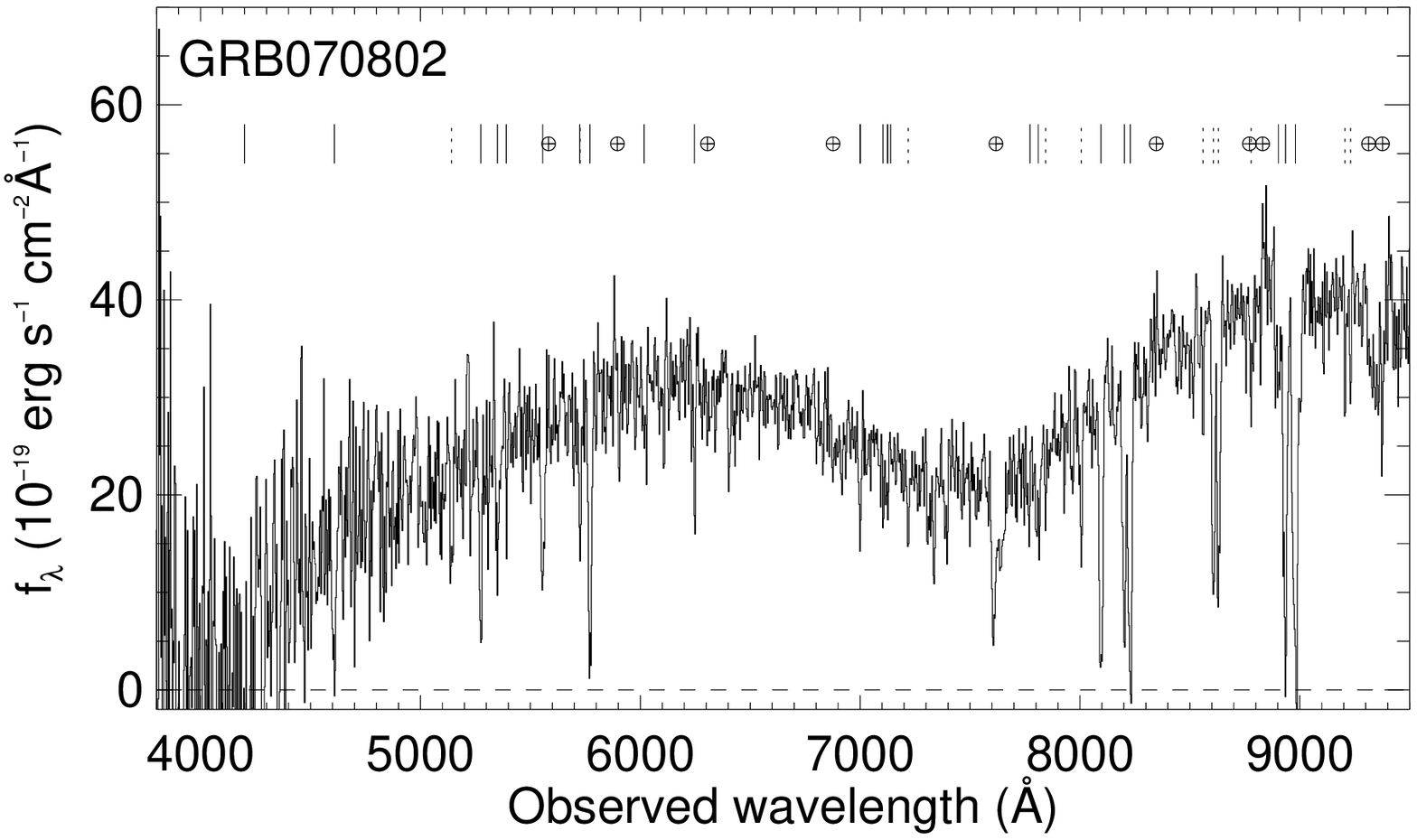}
\caption{
The {\it VLT}\/FORS2 spectrum of the afterglow of GRB\,070802 (El\'iasd\'ottir
et al.\ 2009).
Plotted is the flux-calibrated spectrum
against observed wavelength. Metal lines at the host redshift are
marked with solid lines whereas the features from two intervening systems
are marked with dotted lines. The broad depression centred around 7500 \AA \
is caused by the 2175 \AA \ extinction bump in the host system
at $z_{\rm abs}=2.4549$.
}
\label{bump}
\end{center}
\end{figure}

%\subsection{Metallicities} 
%Afterglow spectroscopy often allows us to measure the
%metallicity of the line-of-sight in the host galaxy. In Fig.~\ref{Z} we plot the
%metallicities along GRB sightlines together with metallicities derived from QSO
%damped Lyman-$\alpha$ absorbers (QSO-DLAs). Here it can be seen that GRBs are
%more metal rich than QSO-DLAs at similar redshifts. Some of the GRB sightlines
%are almost as metal rich as the Lyman-break galaxies at similar redshifts 
%(Pettini et al. 2001). The shift in metallicity relative to QSO-DLAs can be
%understood from the different selection functions of the (star-formation
%selected) GRB-DLAs and the (HI cross-section selected) QSO-DLAs combined
%most likely with metallicity gradients in high-z galaxies (Fynbo et al.
%2008). Hence, most likely GRBs will give a reasonably unbiased census of
%where the massive stars located, at least at $z > 2$ (Fynbo et al. 2006b). 

\subsection{Extinction}
In addition to HI column densities, metal and molecular abundances and
kinematics, the afterglow spectra also provide information about the extinction
curves. The intrinsic spectrum of the afterglow is predicted from theory to be
a power-law and therefore any curvature or other broad features in the spectrum
can be interpreted as being due to the extinction curve shape. So far,
almost all the extinction curves derived for GRB sightlines have
been consistent with an extinction curve similar to that of the SMC
(e.g., Starling et al.\ 2007, Schady et al.\ 2010).  Recently,
a clear detection of the 2175 \AA \ bump known from the
Milky Way was found in a $z=2.45$ GRB sightline (El\'iasd\'ottir et al.\ 2009,
Kr\"uhler et al.\ 2008 and
Fig.~\ref{bump}). This GRB absorber also has unusually strong metal lines
suggesting that the presence of the 2175 {\AA}  extinction bump is related to
high metallicity (as expected from sightlines in the local group). However, we 
have examples of GRBs with nearly solar
metallicity for which the bump is not seen, so it seems that metallicity is not
the only parameter controlling the presence of the 2175 {\AA} extinction bump.
Concerning the amount of extinction, the GRB sightlines vary from no extinction
(e.g., GRB\,050908, Fig.~\ref{LyC}) to $A_V > 5$ mag (e.g., GRB\,070306, Jaunsen et
al.\ 2008).

\subsection{The host sample}\label{sc:sample}

\begin{figure}
\includegraphics[width=\columnwidth]{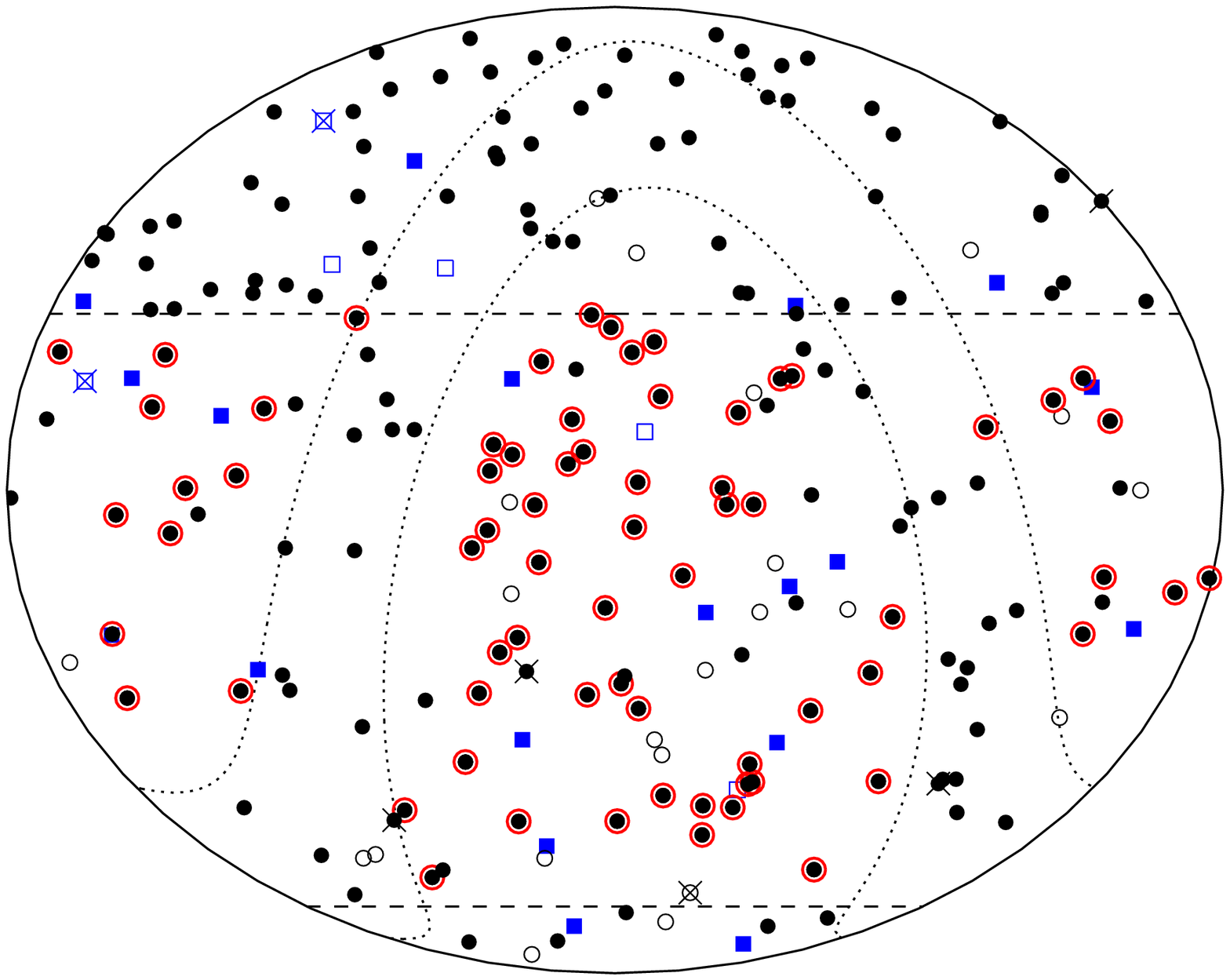} 
\caption{All-sky map (Mollweide projection) of the 238 \textit{Swift} GRBs
occurred between 2005 March and 2007 August. Empty circles: GRBs with no
\textit{Swift}/XRT detection. Filled circles: GRBs with \textit{Swift}/XRT
detection. Filled, encircled circles: GRBs obeying all selection
criteria of the Hjorth et al. (2010) sample. Squares: GRBs classified as short. Crosses: nontriggered
GRBs. Overplotted are the declination cuts ($-70$ and $+27^\circ$, dashed
lines) and  the region with Galactic latitude $|b| > 20^\circ$ (dotted curves),
which roughly corresponds to the sample selection criterion $A_V < 0.5$
mag.\label{fg:map}}
\end{figure}

Hjorth and collaborators have been working on building up a sample of {\it
Swift} GRB host galaxies observed with the ESO {\it VLT}. The main science driver for this
work is to build a representative, unbiased sample of GRB host galaxies than
can be used to firmly establish the statistical properties of GRB hosts. Similar to
the philosophy of the afterglow sample discussed above, this work 
focusses on the systems with the best observability, which also
have the best available information. GRBs included in the sample fullfil the
following criteria: 
\begin{enumerate}
  \item triggered by BAT;
  \item belong to the long-duration class;
  \item trigger time between March 2005 and August 2007;
  \item low Galactic extinction ($A_V \le 0.5$ mag);
  \item prompt XRT localization ($< 12$~hr);
  \item good observability from the {\it VLT} ($-70^\circ < \delta < +27^\circ$);
  \item small position error ($\mbox{radius} \le 2''$) from based on the X-ray 
  afterglow astrometry.
\end{enumerate}
Note that criteria (iii)--(vi) do not introduce selection effects in the sample,
since they are not based on intrinsic properties of the GRBs. Only criterion 
(vii) may in principle bias against faint events (which will have on the average
worse localizations), however in practice it only excludes a few events (3\%
of the total). On the other hand, a burst satisfying all the above criteria
must have been on the average better studied and characterized, since its
afterglow was well observable from the ground.

There are 69 GRBs fulfilling the above criteria. For 52 (75\%) of these the
optical/NIR afterglow has been detected and for 38 (55\%) the redshift has been
measured (spanning the range $0.033 < z < 6.295$).  Figure~\ref{fg:map} shows
the distribution of this sample in the sky.  For more details about the
program, we refer to Hjorth et al.\ (2012).

\subsubsection{Results}\label{sc:results}

\begin{figure}
\includegraphics[width=0.49\columnwidth]{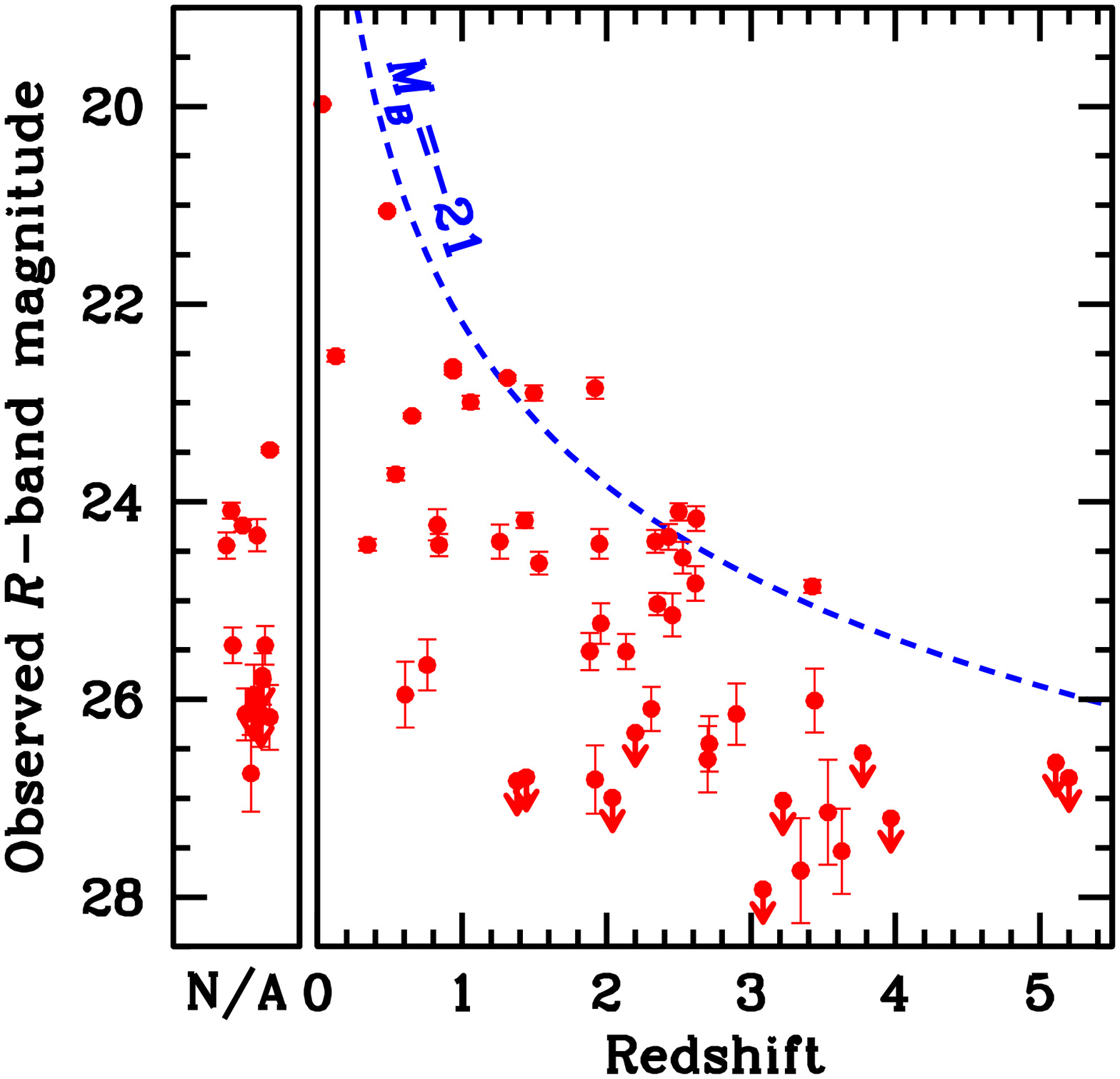}\hspace{0.02\columnwidth}%
\includegraphics[width=0.49\columnwidth]{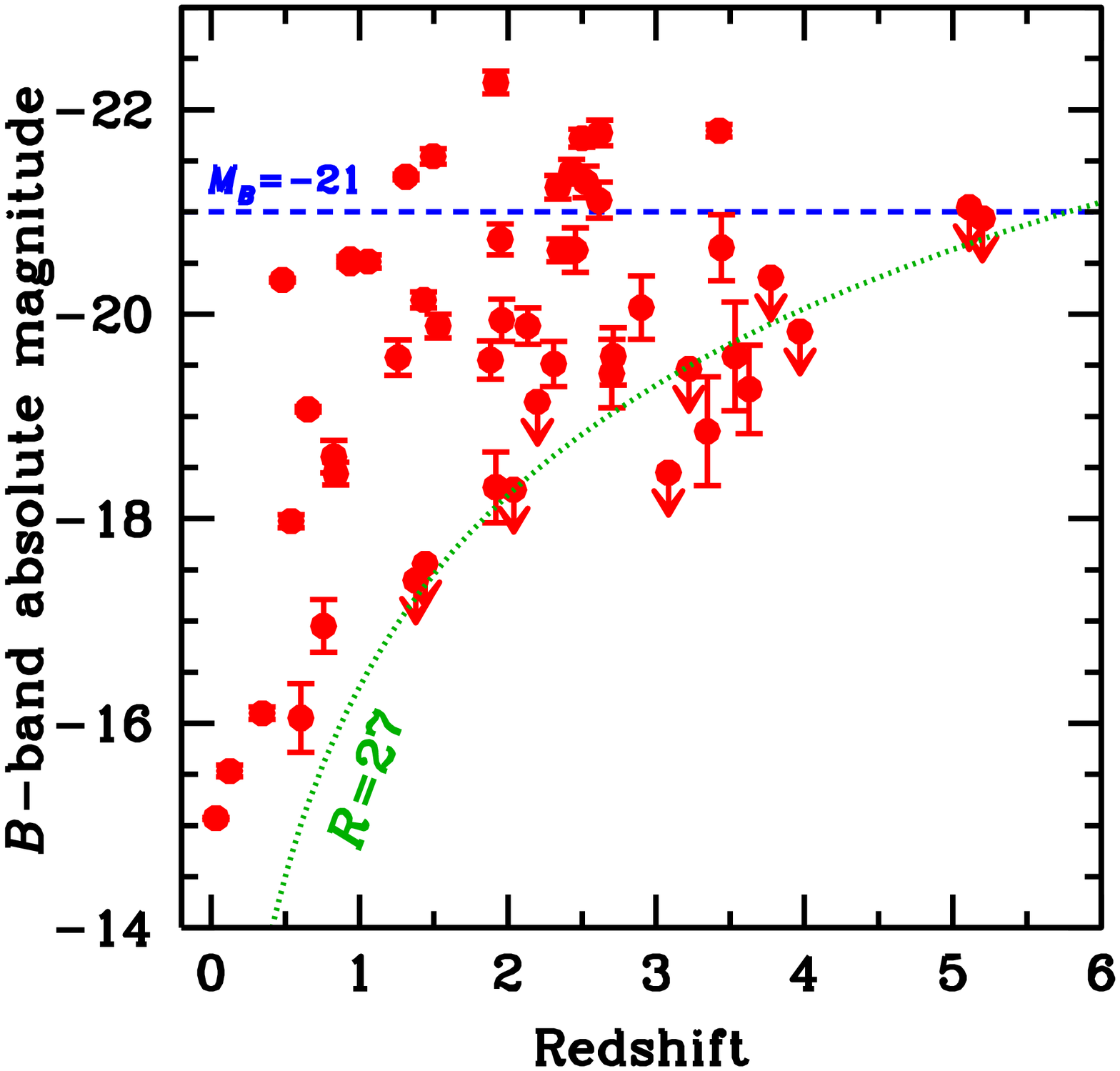}
\caption{Top: observed $R$-band magnitudes of the hosts in the Hjorth et al.\ (2012) sample as a
function of redshift. The left panel shows objects with unknown redshift (the
$x$-axis value is arbitrary). The dashed line shows the magnitude of an $L^*$
galaxy ($M_B = -21$) as a function of redshift. Bottom: absolute luminosity of
hosts as a function of redshift. The dashed line shows the level of $L^*$,
while the dotted curve indicates the effective survey limit ($R \sim
27$).\label{fg:mags}}
\end{figure}

\textbf{Magnitudes.} The primary objective in this study is the search and
localization of the hosts. The applied strategy is a moderately deep $R$-band
exposure (30 min), followed by deeper imaging ($\sim 2$~hr) in case of no
detection. The survey was quite successful, with a detection rate of 80\% of
the hosts.  The success rate drops significantly at large redshifts, with a
recovery fraction of 40\% for GRBs with a measured redshift larger than 3.
Figure~\ref{fg:mags} shows the distribution of the observed and absolute
magnitudes. To compute the $k$-corrections, a spectrum $F_\nu \propto \nu^{-1}$
was assumed. As can be seen, most GRB hosts are subluminous, at a
level $(0.01 \mbox{--} 1) \times L^*$. This is in line with the previous 
findings mentioned above based on smaller and less complete samples (Le Floc'h
et al.\ 2003, Fruchter et al.\ 2006). Hence, this conclusion is not a 
result of a bias against dusty GRB hosts, but it is an intrinsic property 
of the GRB host population.

\begin{figure}
\includegraphics[width=\columnwidth]{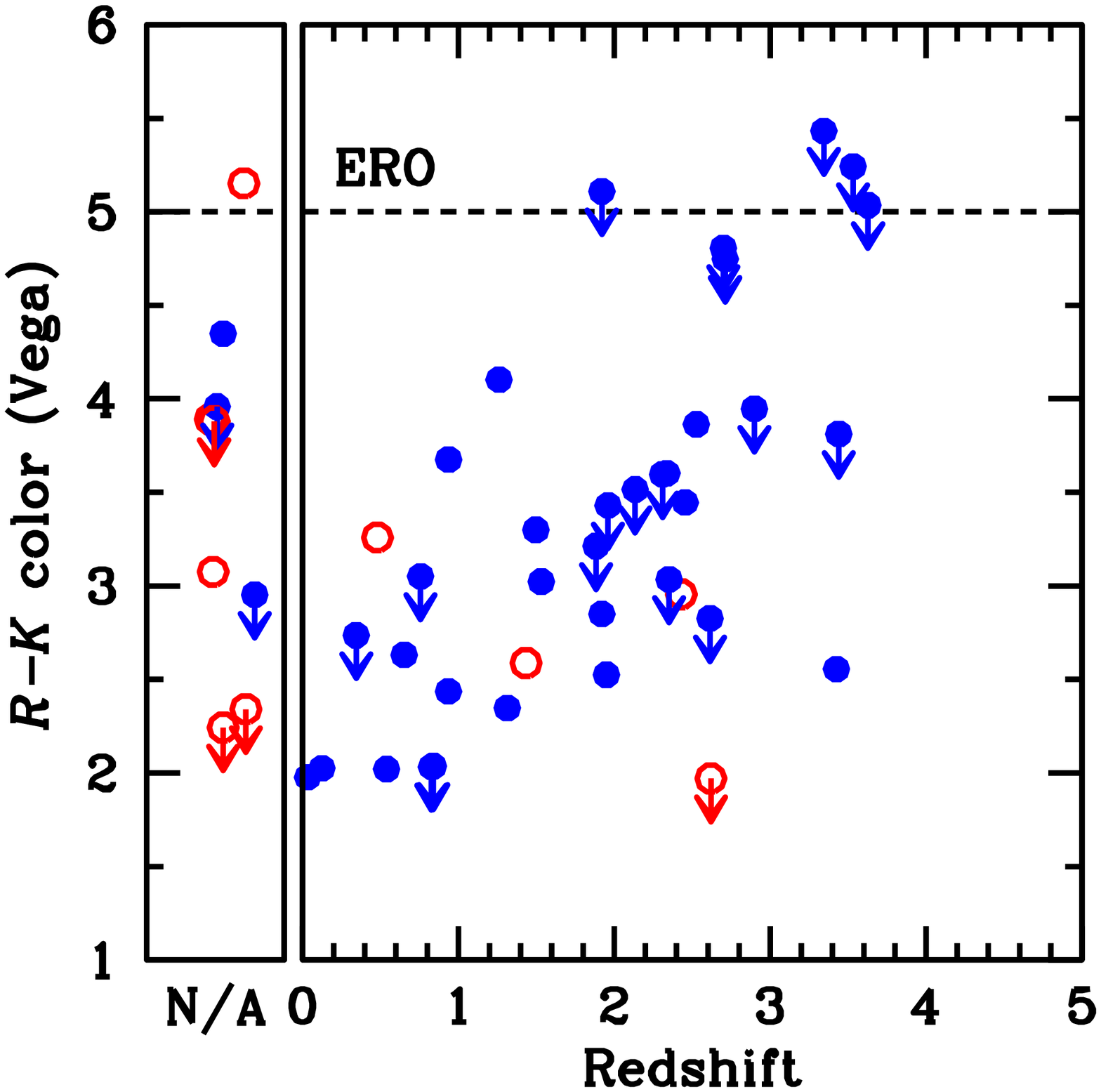}
\caption{$R-K$ colour of the hosts in the Hjorth et al.\ (2012) sample as a function
of $z$. Filled symbols are for GRBs with detected optical afterglows and open
symbols for GRBs with no detected optical afterglow. Only the galaxies with
optical detections are shown. The left panel shows systems with no known
redshift (the $x$-axis value is arbitrary). The horizontal line marks the
boundary of extremely red objects (EROs; $R - K >
5$).\label{fg:colours}}
\end{figure}

\textbf{Colours.} In addition to $R$-band imaging the Hjorth et al.\ (2012) survey also
obtained $K$-band imaging of all the fields. The detection rate at NIR
wavelengths is significantly lower than in the optical.  Hosts were detected in
only about 40\% of the systems in the sample. Overall, colours are in the range $2 <
R-K < 4.5$, with two possible examples of extremely red objects (EROs)
with $R-K \approx 5$. These GRBs had no reported optical afterglow. While the lack of
optical emission is consistent with the presence of dust (and reddening), a
chance association between the GRB and the galaxy cannot be excluded.
Figure~\ref{fg:colours}
shows the distribution of observed colours for bursts with and without an
optical afterglow. Note that the two groups have a comparable distribution of
colours. Overall, even considering bursts with no detected optical afterglow,
the earlier findings of Le Floc'h et al.\ (2003) that GRB hosts have mostly
blue colours are confirmed (though a few cases of red systems have been found
(Levan et al.\ 2006, Berger et al.\ 2007). This does not exclude that dust is
present in these objects (e.g. Jaunsen et al.\ 2008, Tanvir et al.\ 2008), but
is probably confined only in a fraction of the volume occupied by the young
stars (Micha\l{}owski et al.\ 2008).

\textbf{Redshifts.} 
Hosts without known redshift were observed with a variety of
spectroscopic setups (Jakobsson et al.\ 2012, Kr{\"u}hler et al.\ 2012). 
In several cases, only upper and/or lower limits could be placed on the redshift due to the lack of prominent
features. Many GRBs are bound to be in the so-called redshift desert ($1
\lesssim z \lesssim 2$), where the most prominent nebular lines are shifted
into wavelength regions difficult to observe. In some cases, 
this hypothesis was confirmed thanks to the use of red grisms sensitive up to 10,000~\AA,
which probes redshifts up to $z \approx 1.7$ through the [O II] emission line.

\begin{figure}
\includegraphics[width=0.49\columnwidth]{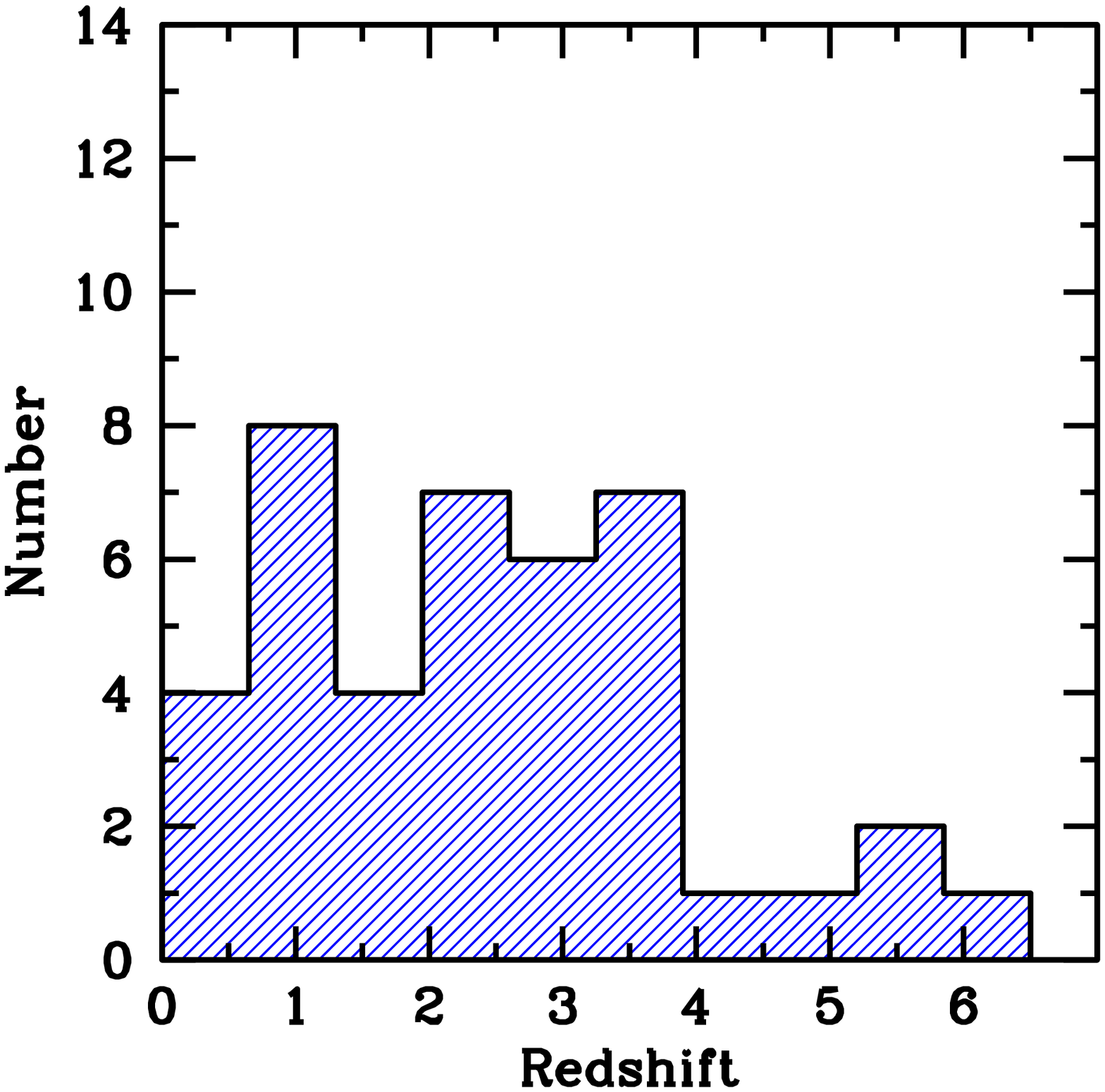}\hspace{0.02\columnwidth}
\includegraphics[width=0.49\columnwidth]{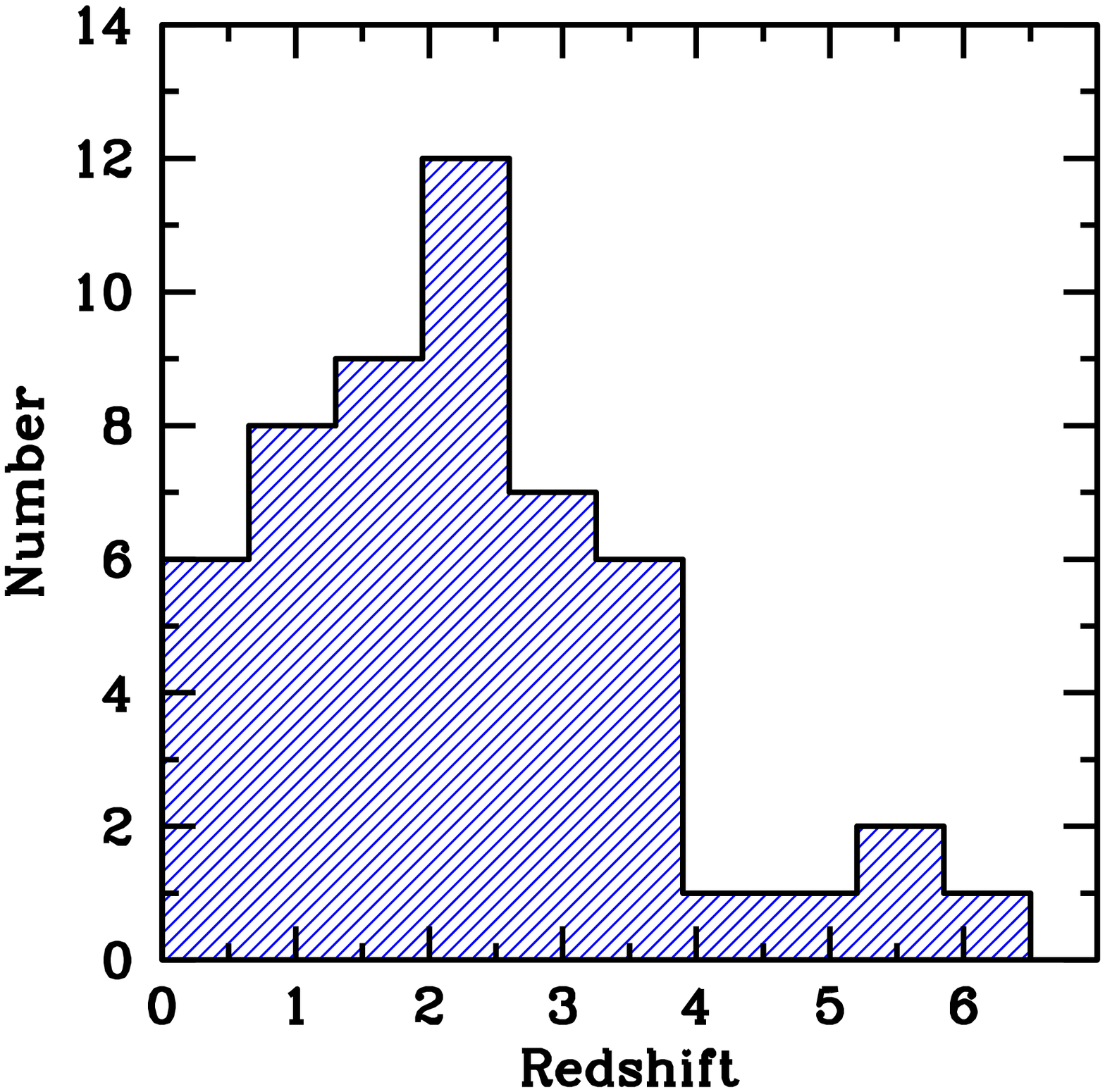}
\caption{Redshift distribution of GRBs in the Hjorth et al.\ (2012) sample. Top
panel: redshifts taken from the literature. Bottom panel: including the results
from host spectroscopy in the program.\label{fg:redshift}}
\end{figure}

Overall, 15 new redshifts were determined from host galaxy spectroscopy, and,
surprisingly, a few of the redshifts reported in the literature were found to
be inconsistent with the redshift derived from the likely host. This is most likely
not due to a wrong host identification as these are bursts with detected
optical afterglows (and for some the reported wrong redshifts were based on
emission lines). For an additional three systems the redshift could be
constrained in the range $1 \lesssim z \lesssim 2$. Surprisingly, only three
of the targeted systems have a redshift larger than 2. While this is partly due
to the selection of the brightest systems for spectroscopy, it also shows that
many dark or optically faint GRBs probably lie at moderate redshifts.
Fig.~\ref{fg:redshift} shows the redshift distribution of the sample, outlining
the contribution from the Hjorth et al.\ (2012) program. We caution that the analysis
of the spectroscopic data is not yet complete. The most noticeable effect is
the reduction of the ``gap'' at $z \sim 1.7$, which was likely due to the lack
of prominent features in the observed spectral range, for both GRB afterglows and hosts (Fiore et al.\ 2007). 

\textbf{Ly$\alpha$ emisson.} For all hosts with a known redshift in the range
$2 < z < 4.5$ (where the Ly$\alpha$ falls in a favorable wavelength range), a
spectrum was obtained with the aim of looking for Ly$\alpha$ in emission
(Milvang-Jensen et al.\ 2012). The
presence of a host galaxy detection in the optical was not required as
Ly$\alpha$ can easily be detected from galaxies that are very faint in the
continuum (e.g., Fynbo et al.\ 2003b).  These spectra also provided a way to
double check some of the redshifts reported in the literature (leading to the
aforementioned discovery of a few likely wrong redshifts). The recovery fraction for these \textit{Swift} GRB
hosts is lower than in earlier cases (with moderately deep exposures of 1.5--4
hr). Ly$\alpha$ is detected in about 35\% of the cases. As mentioned above,
pre-\textit{Swift} studies provided five detections out of five studied cases
(Fynbo et al.\ 2003a). The peak of the Ly$\alpha$ line is observed to be redshifted by a
few hundred km~s$^{-1}$ with respect to the absorption-line redshift inferred
from afterglow spectroscopy. This has been already observed in Lyman break
galaxies (e.g., Adelberger et al.\ 2003).

In summary, the Hjorth et al. (2012) study of {\it Swift} hosts has established that
many of the conclusions reached from the smaller and more biased pre-{\it Swift}
samples still hold: GRB hosts are all star forming and they are predominantly 
(but not all) subluminous and blue.

\section{Simulations of GRB host galaxies}
In this final section we will briefly discuss the work that has been done on
simulating and modeling GRB host galaxies. The modeling of star-forming
galaxies in their cosmological context is a field in rapid progress. Whereas
the treatment of growth of structure in the dark matter component seems to be
very well understood it is still far from trivial to include the baryonic
physics. In particular star formation and its feedback on the interstellar
medium are difficult to include in the simulations. Different ``recipes'' exist,
but it is difficult objectively to establish if they provide an adequate
description of reality rather than providing, e.g., a fitting parameter that
helps reproducing a set of observations. GRB host galaxies are interesting for
testing simulations of galaxies due to their nature as star-formation selected
galaxies. Hence, it is relatively easy to predict from a given simulation what
the properties (e.g., luminosities, metallicities, environments, etc.) of a
sample of GRB host galaxies should be under the assumptions behind the
simulation. This work has only started recently and there is certainly
potential for a lot of development in this area. The first pioneering work
was done by Courty et al.\ (2004, 2007) who basically established that galaxies
similar to the GRB hosts exist in their simulations. Nuza et al.\ (2007)
predicted the properties of GRB hosts under the assumption that GRBs are only
formed by low-metallicity stars. The study used a rather small simulated volume
with a box length of only 10 Mpc, but it was still an important step forward. 
Most lately, Nagamine et al.\ (2008) used a similar simulation to examine the HI absorption
of GRB host absorption systems as a function of redshift.

\section{Conclusions and outlook}
About a decade after the first detection of host galaxies we have progressed
tremendously. GRBs host galaxies have been found to be predominantly 
young, actively star-forming and subluminous. This conclusion seems to 
reflect the intrinsic properties of the host populations and is not only a result of selection effects. The study of the GRB host absorption 
systems is an entire new emerging field providing an interesting link 
between the study of QSO-DLAs and star-forming galaxies selected in 
emission.

There are still unclarified issues. One of the most important questions to 
answer is whether GRBs trace all star-formation or only a limited,
low-metallicity segment. The evidence seems to point in different 
directions. Most likely the road to further progress will be by building
yet more complete samples with more detailed information for all 
GRBs and their hosts (metallicities, luminosities, dust masses, etc.).
Bursts like GRB\,080607 (Prochaska et al.\ 2009) and GRB\,070306 (Jaunsen
et al.\ 2008) suggest that we do preferentially lose GRBs in 
high metallicity, dusty environments. We need to establish the frequency
of such systems in the host population to determine where GRBs fit
into the big picture. Also, more work is needed on improving the predictions
of how the properties of GRB hosts should be under various assumptions 
about the link between GRBs and star-formation.

Another important area for future work is the use of GRBs to probe galaxies at
the epoch of re-ionization. It is now established that GRBs allow us to move
further back in time than what is currently possible with QSOs (Greiner et al.\
2009, Tanvir et al.\ 2010, Salvaterra et al.\ 2010). The discovery of bursts 
like GRB\,090423 also allows us to study in what
type of galaxies most of the star formation happened at $z > 8$, and what was the
nature of the sources responsible for the re-ionization. There is evidence
that the bright $z > 6$ galaxies discovered using colour-colour (drop-out)
selection or more advanced photometric redshifts are too rare to provide the
total star formation rate as well as to have done the re-ionization (e.g.,
Bouwens et al.\ 2007). GRB measurements provide the tool to find the
more typical galaxies responsible for the bulk of the production of ionizing
photons (Ruiz-Velasco et al.\ 2007), and will allow further study of these
galaxies in the future (e.g., with {\it JWST} or 30m ground-based telescopes).

\cleardoublepage
\appendix

\begin{thereferences}{99}
\label{reflist}

\bibitem{adelberger03}
{Adelberger, K.L., et al.} (2003). \textit{ApJ} {\bf 584}, 45.

\bibitem{band98}
{Band, D. L., \& Hartmann, D. H.} (1998). \textit{ApJ} {\bf 493}, 555.

\bibitem{berger07}
{Berger, E., et al.} (2007). \textit{ApJ} {\bf 660}, 504.

\bibitem{berger09}
{Berger, E.} (2009). \textit{ApJ} {\bf 690}, 231.

\bibitem{bloom02}
{Bloom, J. S., Kulkarni, S. R., \& Djorgovski, S. G.} (2002). \textit{AJ} {\bf 123}, 1111.

\bibitem{Bornancini04}
{Bornancini, C. G., et al.} (2004). \textit{ApJ} {\bf 614}, 84.

\bibitem{Bouwens}
{Bouwens, R. J., et al.} (2007). \textit{ApJ} {\bf 670}, 928.

\bibitem{calura}
{Calura, F., et al.} (2009). \textit{ApJ} {\bf 693}, 1236.

\bibitem{campisi08}
{Campisi, M. A. \& Li, L.-X.} (2008). \textit{MNRAS} {\bf 391}, 935. 

\bibitem{castrotirado08}
{Castro-Tirado, A. J., et al.} (2008). \textit{Nat} {\bf 455}, 506.

\bibitem{charlot93}
{Charlot, S. \& Fall, S. M.} (1993). \textit{ApJ} {\bf 415}, 580.

\bibitem{chen05}
{Chen, H.-W., Prochaska, J. X., Bloom, J. S., \& Thompson, I. B.} (2005). \textit{ApJ} {\bf 634}, L25.

\bibitem{chen07}
{Chen, H.-W., et al.} (2007). \textit{ApJL} {\bf 667}, L125.

\bibitem{christensen04} 
{Christensen, L., Hjorth, J., \& Gorosabel, J.} (2004). \textit{A\&A} {\bf 425}, 913.

\bibitem{cobb08}
{Cobb, B. E. \& Baylin, C. D.} (2008). \textit{ApJ} {\bf 677}, 1157.

\bibitem{costa97}
{Costa, E., et al.} (1997). \textit{Nat} {\bf 387}, 783.

\bibitem{courty04}
{Courty, S., Bj\"ornsson, G., \& Gudmundsson, E. H.} (2004). \textit{MNRAS} {\bf 354}, 581.

\bibitem{courty07}
{Courty, S., Bj\"ornsson, G., \& Gudmundsson, E. H.} (2007). \textit{MNRAS} {\bf 376}, 1375.

\bibitem{crowther07}
{Crowther, P.} (2007). \textit{ARA\&A} {\bf 45}, 177.

\bibitem{delia07}
{D'Elia, V., et al.} (2007). \textit{A\&A} {\bf 467}, 629.

\bibitem{djorgovski}
{Djorgovski, G., et al.} (2003). {\it Proc. SPIE} (Edited by Guhathakurta, Puragra) {\bf 4834}, 238. 

\bibitem{eliasdottir}
{El\'iasd\'ottir, \'A., et al.} (2009). \textit{ApJ} {\bf 697}, 1725.

\bibitem{fenimore93}
{Fenimore, E. E., et al.} (1993). \textit{Nature} {\bf 366}, 40.

\bibitem{fiore07}
{Fiore, F., et al.} (2007). \textit{A\&A} {\bf 470}, 515.

\bibitem{Foley06}
{Foley, S., et al} (2006). \textit{A\&A} {\bf 447}, 891.

\bibitem{fruchter06}
{Fruchter, A. S., et al.} (2006). \textit{Nat} {\bf 441}, 463.

\bibitem{fynbo01}
{Fynbo, J. P. U., et al.} (2001). \textit{A\&A} {\bf 369}, 373.

\bibitem{fynbo02}
{Fynbo, J. P. U., et al.} (2002). \textit{A\&A} {\bf 388}, 425.

\bibitem{fynbo03a}
{Fynbo, J. P. U., et al.} (2003a). \textit{A\&AL} {\bf 406}, L63.

\bibitem{fynbo03b}
{Fynbo, J. P. U., et al.} (2003b). \textit{A\&A} {\bf 407}, 147.

\bibitem{fynbo05}
{Fynbo, J. P. U., et al.} (2005) \textit{ApJ} {\bf 633}, 317.

\bibitem{fynbo06b}
{Fynbo, J. P. U., et al.} (2006). \textit{A\&AL} {\bf 451}, L47.

\bibitem{fynbo08}
{Fynbo, J. P. U., et al.} (2008). \textit{ApJ} {\bf 683}, 321.

\bibitem{fynbo09}
{Fynbo, J. P. U., et al.} (2009). \textit{ApJS} {\bf 185}, 526.

\bibitem{galama98}
{Galama, T. J., et al.} (1998). \textit{Nat} {\bf 395}, 670.

\bibitem{gehrels04}
{Gehrels, N., et al.} (2004). \textit{ApJ} {\bf 611}, 1005. 

\bibitem{greiner}
{Greiner, J., et al.} (2009). \textit{ApJ} {\bf 693}, 1610.

\bibitem{groot98}
{Groot, P., et al.} (1998). \textit{ApJL} {\bf 493}, L27. 

\bibitem{Hjorth12}
{Hjorth, J., et al.} (2012). \textit{ApJ}, {\bf 756}, 187. 

\bibitem{hogg99}
{Hogg, D. W. \& Fruchter, A. S.} (1999). \textit{ApJ} {\bf 520}, 54.

\bibitem{hurley93}
{Hurley, K., et al.} (1993). \textit{A\&AS} {\bf 97}, 39.

\bibitem{palli04a}
{Jakobsson, P., et al.} (2004a). \textit{ApJL} {\bf 617}, L21.

\bibitem{palli04b}
{Jakobsson, P., et al.} (2004b). \textit{A\&A} {\bf 427}, 785.

\bibitem{palli05a}
{Jakobsson, P., et al.} (2005a). \textit{ApJ} {\bf 629}, 45.

\bibitem{palli05b}
{Jakobsson, P., et al.} (2005b) \textit{MNRAS} {\bf 362}, 245.

\bibitem{palli06a}
{Jakobsson, P., et al.} (2006a). \textit{A\&A} {\bf 447}, 897.

\bibitem{palli06b}
{Jakobsson, P., et al.} (2006b). \textit{A\&AL} {\bf 460}, L13.

\bibitem{palli12}
{Jakobsson, P., et al.} (2012). \textit{ApJ} {\bf 752}, 62.

\bibitem{jaunsen08}
{Jaunsen, A. O., et al.} (2008). \textit{ApJ} {\bf 681}, 453.

\bibitem{kasliwal08}
{Kasliwal, M. M., et al.} (2008). \textit{ApJ} {\bf 678}, 1127.

\bibitem{kelly07}
{Kelly, P. L., Kirshner, R. P., \& Pahre, M.} (2007). \textit{ApJ} {\bf 687}, 1201.

\bibitem{kruehler08}
{Kr\"uhler, T., et al.} (2008). \textit{ApJ} {\bf 685}, 376.

\bibitem{kruehler12}
{Kr\"uhler, T., et al.} (2012). \textit{ApJ} {\bf 758}, 46.

\bibitem{kullarni98}
{Kulkarni, S., et al.} (1998). \textit{Nature} {\bf 393}, 35.

\bibitem{kurk00}
{Kurk, J., et al.} (2000). \textit{A\&AL} {\bf 358}, 1.

\bibitem{larson97}
{Larson, S. B.} (1997). \textit{ApJ} {\bf 491}, 86.

\bibitem{larsson07}
{Larsson, J., Levan, A. J., Davies, M. D., \& Fruchter, A. S.} (2007). \textit{MNRAS} {\bf 376}, 1285.

\bibitem{lazzati02}
{Lazzati, D., Covino, S., \& Chisellini, G.} (2002). \textit{MNRAS} {\bf 330}, 583.

\bibitem{lefloch03}
{Le Floc'h, E., et al.} (2003). \textit{A\&AL} {\bf 400}, L499.

\bibitem{levan06}
{Levan, A., et al.} (2006). \textit{ApJ} {\bf 647}, 471.

\bibitem{mao98}
{Mao, S. \& Mo, H.~J.} (1998). \textit{ApJL} {\bf 339}, L1.

\bibitem{mas-hesse}
{Mas-Hesse, J. M., et al.} (2003). \textit{ApJ} {\bf 598}, 858.

\bibitem{metzger97}
{Metzger, M. R., et al.} (1997). \textit{Nature} {\bf 387}, 878.

\bibitem{michal08}
{Micha\l{}owski, M., Hjorth, J., Castro-Cer\'on, J. M., \& Watson, D.} (2008). \textit{ApJ} {\bf 672}, 817.

\bibitem{Bo12}
{Milvang-Jensen, B., et al.} (2012). \textit{ApJ} {\bf 756}, 25.

\bibitem{modjaz08}
{Modjaz, M., et al.} (2008). \textit{AJ} {\bf 135}, 1136.

\bibitem{nagamine08}
{Nagamine, K., Zhang, B., \& Hernquist, L.} (2008). \textit{ApJ} {\bf 686}, L57.

\bibitem{natarajan97}
{Natarajan, P., et al.} (1997). \textit{NewA} {\bf 2}, 471. 

\bibitem{nuza}
{Nuza, S. E., et al.} (2007). \textit{MNRAS} {\bf 375}, 665.

\bibitem{paczynski98}
{Paczy\'ynski, B.} (1998). \textit{ApJL} {\bf 494}, L45.

\bibitem{pattel10}
{Pattel, M., et al.} (2010) \textit{A\&AL} {\bf 512}, L3.

\bibitem{pedersen06}
{Pedersen, K., et al.} (2006). \textit{ApJ} {\bf 636}, 381. 

\bibitem{priddey06}
{Priddey, R. S., et al.} (2006). \textit{MNRAS} {\bf 369}, 1189.

\bibitem{pgw+03}
{Prochaska, J.~X.,et al.} (2003). \textit{ApJL} {\bf 595}, L9.

\bibitem{prochaska07}
{Prochaska, J. X., Chen, H.-W., Dessauges-Zavadsky, M., \& Bloom, J. S.} (2007). \textit{ApJ} {\bf 666}, 267.

\bibitem{prochaska09}
{Prochaska, J. X., et al.} (2009). \textit{ApJ} {\bf 691}, L27.

\bibitem{rol05}
{Rol, E., et al.} (2005). \textit{ApJ} {\bf 624}, 868. 

\bibitem{rol07}
{Rol, E., et al.} (2007). \textit{ApJ} {\bf 669}, 1098.

\bibitem{ruiz-velasco07}
{Ruiz-Velasco, A. E., et al.} (2007). \textit{ApJ} {\bf 669}, 1.

\bibitem{salva10}
{Salvaterra, R. et al.} (2010). \textit{Nature} {\bf 461}, 1258.

\bibitem{sahu97}
{Sahu, K. S., et al.} (1997). \textit{Nature} {\bf 387}, 476.

\bibitem{savaglio06}
{Savaglio, S.} (2006). \textit{NJP} {\bf 8}, 195.

\bibitem{schady2010}
{Schady, P. et al.} (2010). \textit{MNRAS} {\bf 401}, 2773.

\bibitem{silva98}
{Silva, L., et al.} (1998). \textit{ApJ} {\bf 509}, 103. 

\bibitem{soderberg04}
{Soderberg, A., Djorgovski, S. G., Halpern, J. P., \& Mirabal, N.} (2004). \textit{GCN} {\bf 2837}, 1.

\bibitem{starling05}
{Starling, R.L.C., et al.} (2005). \textit{A\&A} {\bf 442}, L21.

\bibitem{starling07}
{Starling, R.L.C., et al.} (2007). \textit{ApJ} {\bf 661}, 787.

\bibitem{tanvir04}
{Tanvir, N., et al.} (2004). \textit{MNRAS} {\bf 352}, 1073.

\bibitem{tanvir08}
{Tanvir, N., et al.} (2008). \textit{MNRAS} {\bf 388}, 1743.

\bibitem{tanvir10}
{Tanvir, N. et al.} (2010). \textit{Nature} {\bf 461}, 1254.

\bibitem{vanParadijs97}
{van Paradijs, J., et al. } (1997). \textit{Nat} {\bf 386}, 686. 

\bibitem{vanparadijs00}
{van Paradijs, J., Kouveliotou, C., \& Wijers, R. A. M. J. } (2000).  \textit{ARAA} {\bf 38}, 379.

\bibitem{vreeswijk04}
{Vreeswijk, P. M., et al.} (2004). \textit{A\&A} {\bf 419}, 927.

\bibitem{vreeswijk07}
{Vreeswijk, P. M., et al.} (2007). \textit{A\&A} {\bf 468}, 83.

\bibitem{wijers98}
{Wijers, R. A. M. J., Bloom, J. S., Bagla, J. S., \& Natarajan, P.} (1998). \textit{MNRAS Letters} {\bf 294}, L13.

\bibitem{wijnand09}
{Wijnands, R. et al.} (2009). \textit{MNRAS} {\bf 393}, 126.

\bibitem{wolf07}
{Wolf, C. \& Podsiadlowski, P.} (2007). \textit{MNRAS} {\bf 375}, 1049.

\bibitem{Wolfe05}
{Wolfe, A. M., Gawiser, E., \& Prochaska, J. X.} (2005). \textit{ARA\&A} {\bf 43}, 861.

\bibitem{woods95}
{Woods, E. \& Loeb, A.} (1995).\textit{ApJ} {\bf 453}, 583.

\end{thereferences}
\end{document}